\documentclass[debug]{rmaa}

\usepackage[dvipsnames]{xcolor}
\usepackage{graphicx}
\usepackage{xspace}
\usepackage{txfonts}
\usepackage{multirow}
\usepackage{lscape}
\usepackage{natbib,twoopt}
\usepackage{color}% Nueva línea de texto. 
\usepackage{soul}
\usepackage{longtable}
\usepackage{geometry}
\usepackage{subfig}
\usepackage{tabularx}
\usepackage[breaklinks=true]{hyperref} % to avoid \citeads line fills
\bibpunct{(}{)}{;}{a}{}{,} % to follow the A&A style

\makeatletter
\newcommandtwoopt{\citeads}[3][][]{\href{http://adsabs.harvard.edu/abs/#3}%
{\def\hyper@linkstart##1##2{}%
\let\hyper@linkend\@empty\citealp[#1][#2]{#3}}}
\newcommandtwoopt{\citepads}[3][][]{\href{http://adsabs.harvard.edu/abs/#3}%
{\def\hyper@linkstart##1##2{}%
\let\hyper@linkend\@empty\citep[#1][#2]{#3}}}
\newcommandtwoopt{\citetads}[3][][]{\href{http://adsabs.harvard.edu/abs/#3}%
{\def\hyper@linkstart##1##2{}%
\let\hyper@linkend\@empty\citet[#1][#2]{#3}}}
\newcommandtwoopt{\citeyearads}[3][][]%
{\href{http://adsabs.harvard.edu/abs/#3}
{\def\hyper@linkstart##1##2{}%
\let\hyper@linkend\@empty\citeyear[#1][#2]{#3}}}
\makeatother

\newcommand{\ergs}{\ensuremath{\mathrm{erg\,s}^{-1}}\xspace}

\newcommand{\ergscm}{\ensuremath{\mathrm{erg\,s}^{-1}\mathrm{cm}^{-2}}\xspace}

\newcommand{\xmm}{\textsl{XMM-Newton}\xspace}

\newcommand{\chandra}{\textsl{Chandra}\xspace}

\newcommand{\suzaku}{\textsl{Suzaku}\xspace}

\newcommand{\asca}{\textsl{ASCA}\xspace}
\newcommand{\maxi}{\textsl{MAXI}/GSC\xspace}
\newcommand{\fermi}{\textsl{Fermi}/GBM\xspace}

\newcommand{\centau}{Cen~X-3\xspace}
\newcommand{\gx}{GX~301-2\xspace}

\newcommand{\hmxbs}{HMXBs\xspace}

\usepackage{paralist}

\usepackage{psfrag,color}

\usepackage[latin1]{inputenc}

\title{\gx pre-periastron and apastron flares with \emph{MAXI}} 

\author{
  \'A. Torregrosa\altaffilmark{{2}}, 
  J. J. Rodes-Roca\altaffilmark{1,2},
J. M. Torrej\'on\altaffilmark{1,2},
G. Sanjurjo-Ferr\'in\altaffilmark{{2}},
T. Mihara\altaffilmark{{3}},
M. Nakajima\altaffilmark{{4}},
and M. Sugizaki\altaffilmark{{5}}}

\altaffiltext{1}{Department of Physics, Systems Engineering and Signal Theory, University of Alicante,
              03080 Alicante, Spain.}

\altaffiltext{2}{University Institute of Physics Applied to Sciences and Technologies,
              University of Alicante, 03080 Alicante, Spain.}

\altaffiltext{3}{MAXI team, Institute of Physical and Chemical Researh (RIKEN), 2-1 Hirosawa, Wako, Saitama 351-0198, Japan.}

\altaffiltext{4}{Nihon University School of Dentistry at Matsudo 2-870-1 Sakae-cho Nishi, Matsudo-shi, Chiba 271-8587, Japan.}

\altaffiltext{5}{Science and Engineering, Kanazawa University, Kakuma, Kanazawa, Ishikawa 920-1192, Japan.}

\shortauthor{\'A. Torregrosa et al.}
\shorttitle{\gx pre-periastron and apastron flares with \emph{MAXI}}

\fulladdresses{
\item \'A. Torregrosa, J. J. Rodes-Roca, J. M. Torrej\'on and G. Sanjurjo-Ferr\'in: University Institute of Physics Applied to Sciences and Technologies, University of Alicante, 03080 Alicante, Spain (aat34@alu.ua.es).
\item J. J. Rodes-Roca and J. M. Torrej\'on: Department of Physics, Systems Engineering and Signal Theory, University of Alicante, 03080 Alicante, Spain.
\item T. Mihara: MAXI team, Institute of Physical and Chemical Researh (RIKEN), 2-1 Hirosawa, Wako, Saitama 351-0198, Japan.
\item M. Nakajima: Nihon University School of Dentistry at Matsudo 2-870-1 Sakae-cho Nishi, Matsudo-shi, Chiba 271-8587, Japan.
\item M. Sugizaki: Science and Engineering, Kanazawa University, Kakuma, Kanazawa, Ishikawa 920-1192, Japan.
}

\listofauthors{\'A. Torregrosa, J. J. Rodes-Roca, J. M. Torrej\'on, G. Sanjurjo-Ferr\'in, T. Mihara, N. Motoki, M. Sugizaki}
\indexauthor{Torregrosa, \'A.}
\indexauthor{Rodes-Roca, J. J.}
\indexauthor{Torrej\'on, J. M.}
\indexauthor{Sanjurjo-Ferr\'in, G.}
\indexauthor{Mihara, T.}
\indexauthor{Nakajima, M.}
\indexauthor{Sugizaki, M.}

\abstract{The bright high-mass X-ray binary \gx exhibits two periodic flare episodes along its orbit which are produced when the neutron star is close to the apastron and periastron passages. Time-resolved spectra were extracted and several models applied to describe all of them.  The best description was obtained with a blackbody continuum modified by the Fe K-shell absorption edge, absorbed by a large column density on the order of $10^{23}$ cm$^{-2}$ and, if present, a Fe K$\alpha$ fluorescent emission line. Three of the nine apastron flares were as bright as the pre-periastron flare and two of them coincided with spin-up episodes of the neutron star. This fact points to the presence of a transient disc around the neutron star as it passes through the apastron that increases the accretion process. The size of emitting region on the neutron star surface showed some variability but quite consistent with a hot spot.

}

\resumen{\gx presenta dos fulguraciones peri\'odicas que se producen cuando la estrella de neutrones se encuentra cerca de los pasos por el apoastro y el periastro. Se extrajeron los espectros con resoluci\'on en fase orbital y la mejor descripci\'on de todos ellos se obtuvo con un continuo de cuerpo negro modificado por el umbral de absorci\'on de la capa K del Fe, absorbido por una gran columna de hidr\'ogeno ($10^{23}$ cm$^{-2}$) y, si estaba presente, una l\'inea de emisi\'on fluorescente del Fe K$\alpha$. Tres de las nueve fulguraciones en el apoastro fueron tan brillantes como la fulguraci\'on del preperiastro y dos de ellos coincidieron con episodios de aumento del per\'iodo de la pulsaci\'on de la estrella de neutrones. Este hecho apunta a la presencia de un disco transitorio alrededor del objeto compacto en su paso por el apoastro. El tama\~no de la regi\'on de emisi\'on se mostr\'o consistente con una mancha caliente en la superficie del objeto compacto.
}

\addkeyword{pulsars: individual: GX 301--2}
\addkeyword{stars: supergiants}
\addkeyword{X-rays: binaries}

\begin{document}
% Typeset article header
\maketitle

\section{Introduction}
\label{introduction}

GX 301$-$2 (also known as 4U 1223$-$62) is an X-ray pulsar whose rapid variability in the high-energy X-ray flux was discovered by \citet{1971ApJ...166L..73M} and \citet{1971ApJ...166L..69L}. The neutron star moves in an eccentric orbit ($e = 0.462$) with an orbital period of 41.5 days \citep{1986ApJ...304..241S} around the early B-type companion, Wray 15-977 (BP Cru). Its spin period of $\sim$ 685 s was  discovered by \citet{1976ApJ...209L.119W} and long-term frequency history shows a complex behaviour spinning up and down on different timescales, as it can be seen in the \emph{Fermi}/GBM Accreting Pulsars Program \citep{2009arXiv0912.3847F,2020ApJ...896...90M}. The X-ray light curve shows a regular strong peak which is due to the pre-periastron passage of the neutron star. At this orbital phase, the compact object not only accretes matter from the dense stellar wind but is further enhanced by the material in the circumstellar disc around the compact object, which in turns powers the X-ray emission. Besides this strong flare, evidence of a second though much weaker periodic flare near apastron passage was discovered by \citet{1995ApJ...454..872P}. However, it is not always detectable \citep{2001ApJ...554..383P}. Spherically symmetric stellar wind models fail to described both column density and luminosity data. Nevertheless, adding a gas stream which is not spherically distributed around the optical companion, improved the description of the data producing strong peaks in both column density and luminosity data \citep{1988MNRAS.232..199S,1991MNRAS.250..310L,1991A&A...252..272H}.

The spectral classification of Wray 15-977 is B1 Ia+ hypergiant \citep{1995A&A...300..446K}, confirmed by the spectra taken with the high-resolution Ultraviolet and Visual Echelle Spectrograph on the Very Large Telescope \citep{2006A&A...457..595K}. Model atmosphere fits to high-resolution optical spectra gave a radius $R_\star = 62\, R_\sun$, a mass $M_\star = \left(43^{+10}_{-4}\right)$ $M_\sun$, an effective temperature $T_\mathrm{eff} = 18\,000$ K and  a luminosity  $L_\star = 5\times 10^5\, L_\sun$. This luminosity corresponds to a distance $D = 3$ kpc, proposed by \citet{2006A&A...457..595K} on the basis of the \ion{Na}{i} D interstellar lines. The European Space Agency (ESA) mission Gaia\footnote{https://www.cosmos.esa.int/gaia} Early Data Release 3 (GEDR3) has been used to find the distance to \gx, $D = \left(3.55^{+0.18}_{-0.16}\right)$ kpc. It is known as ''photogeometric'' distance and has been determined by using the parallax measure, BR-RP colour and the source's G-band magnitude \citep{2021AJ....161..147B}. %Bailer-Jones et al. (2021)

GX 301--2 has been observed by most X-ray observatories such as \emph{Tenma} \citep{1991MNRAS.250..310L}, \emph{EXOSAT} \citep{1991A&A...252..272H}, \asca \citep{1996ApJ...463..726S,2002ApJ...574..879E}, \emph{RXTE} \citep{2004A&A...427..567M}, \emph{BeppoSAX} \citep{2005A&A...438..617L}, \chandra \citep{2003ApJ...597L..37W} and \suzaku \citep{2012ApJ...745..124S}. Focusing the study of the spectrum in the [0.3--10.0] keV energy band, the X-ray continuum spectra are usually described by three components: (i) an extremely absorbed power law; (ii) a scattered high absorbed power law; and (iii) an absorbed thermal component with a temperature of 0.8 keV. Moreover, the residuals between the spectrum and the continuum model show a number of emission lines depending on the X-ray observatory and the orbital phase of the observation. In addition, the iron complex at 6.4 keV was fully resolved by \emph{Chandra} detecting the Compton shoulder of the iron K$\alpha$ line at 6.24 keV \citep{2003ApJ...597L..37W}.

The source GX 301$-$2 was observed with the \emph{XMM-Newton} observatory on August 14th, 2008 (MJD 54\,692.11--54\,692.88, ObsID 0555200301), and on July 12th, 2009 (MJD 55\,024.103--55\,024.643, ObsID 0555200401). Both observations were taken during the pre-periastron flare. As a result of the high flux at this orbital phase $\sim 0.91$, the \emph{EPIC/MOS} CCD cameras were turned off to provide more telemetry for the \emph{EPIC/PN} instrument, which was operated in modified timing mode, and medium filter was used. In this timing mode the lower energy threshold of the instrument is increased to 2.8 keV to avoid telemetry drop outs due to the brightness of the source. The orbital-phase averaged spectrum of both observations were extracted and described with several models.
After trying different models, we found that the best continuum model that could fit both spectra significantly well was a hybrid model, combining thermal and non-thermal components. A component is called thermal when radiation is produced as a consequence of the thermal motion of the plasma particles (for instance, blackbody radiation). Otherwise, the emitted radiation is non-thermal (for instance, non-thermal inverse-Compton emission) \citep{2015A&A...576A.108G}. The [2.8--10.0] keV energy spectrum was fitted using a model of the form:

\begin{equation}
  F(E) = tbnew \times po + tbnew \times bbody + GL\, ,
  \label{ec1}
\end{equation}

where \emph{tbnew}\footnote{https://pulsar.sternwarte.uni-erlangen.de/wilms/research/tbabs/}, in terms of \textsc{XSPEC}, is a new version of the T\"{u}bingen-Boulder interstellar medium absorption model which updates the absorption cross sections and abundances \citep{2000ApJ...542..914W}; %Wilms et al.
\emph{po} is a typical photon power law whose parameters include a dimensionless photon index ($\Gamma$) and the normalisation constant (\emph{K}), the spectral photons keV$^{-1}$ cm$^{-2}$ s$^{-1}$ at 1 keV; \emph{bbody} corresponds to a simple blackbody model incorporating the temperature $kT_\mathrm{bb}$ in keV and the normalisation \emph{norm}, defined as $L_{39}/D^{2}_{10}$, where $L_{39}$ is the source luminosity in units of 10$^{39}$ \ergs and $D^{2}_{10}$ is the distance to the source in units of 10 kpc; and \emph{GL} indicates the Gaussian functions added to describe the emission lines. Therefore, the power-law component is interpreted in terms of inverse Compton-scattering by hot electrons of a seed radiation field, while the blackbody radiation is a foot-print of hot-spots on the NS surface. The total absorption column comprises the contribution from the interstellar medium and from the medium local to the source. The latter is by far the most dominant, and can be used to probe stellar wind structures or the presence of matter surrounding the compact object (i.e. accretion stream, accretion disc).

\begin{figure}[hbtp]
  \centering
  \includegraphics[angle=-90.0,width=\columnwidth]{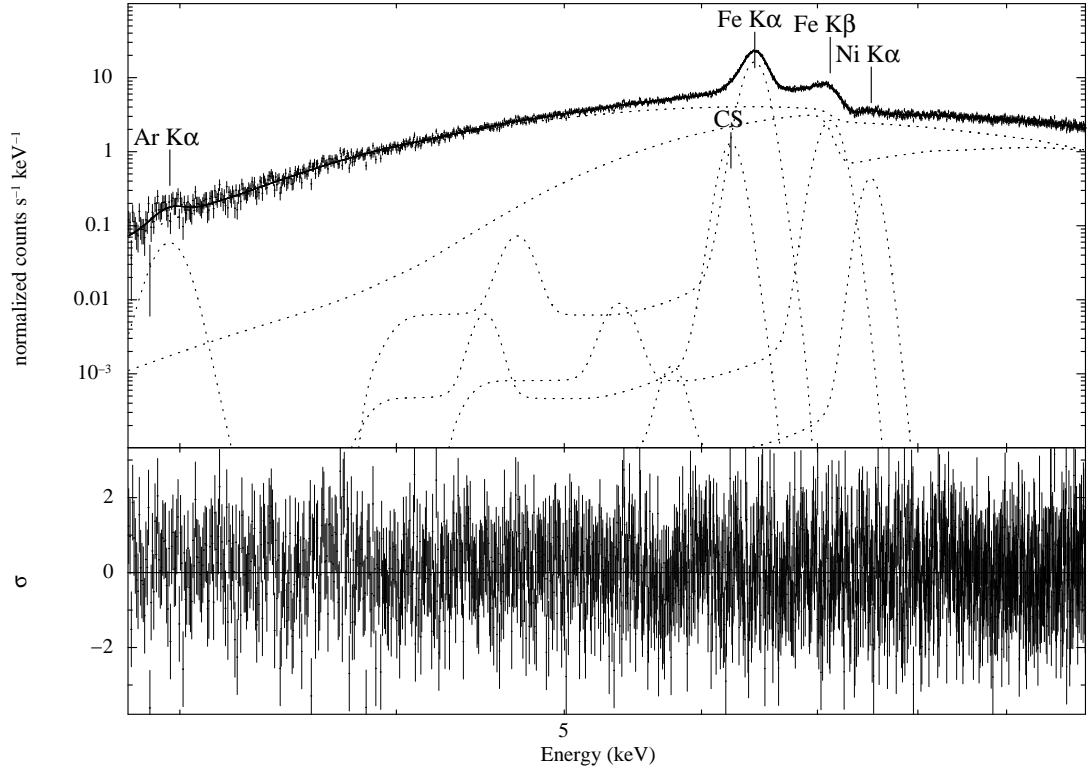}  
  \caption[]{\label{figure1} %
    \xmm with ID 0555200301 (2008) orbital phase averaged spectrum of \gx in the (2.8--10.0) keV band. \emph{Top panel}: data and best-fit model described by equation~(\ref{ec1}). \emph{Bottom panel} shows the residuals between the spectrum and the model. Spectral parameters can be found in Table~\ref{tab:4} (third column).}
\end{figure}

\begin{figure}[hbtp]
  \centering
  \includegraphics[angle=-90.0,width=\columnwidth]{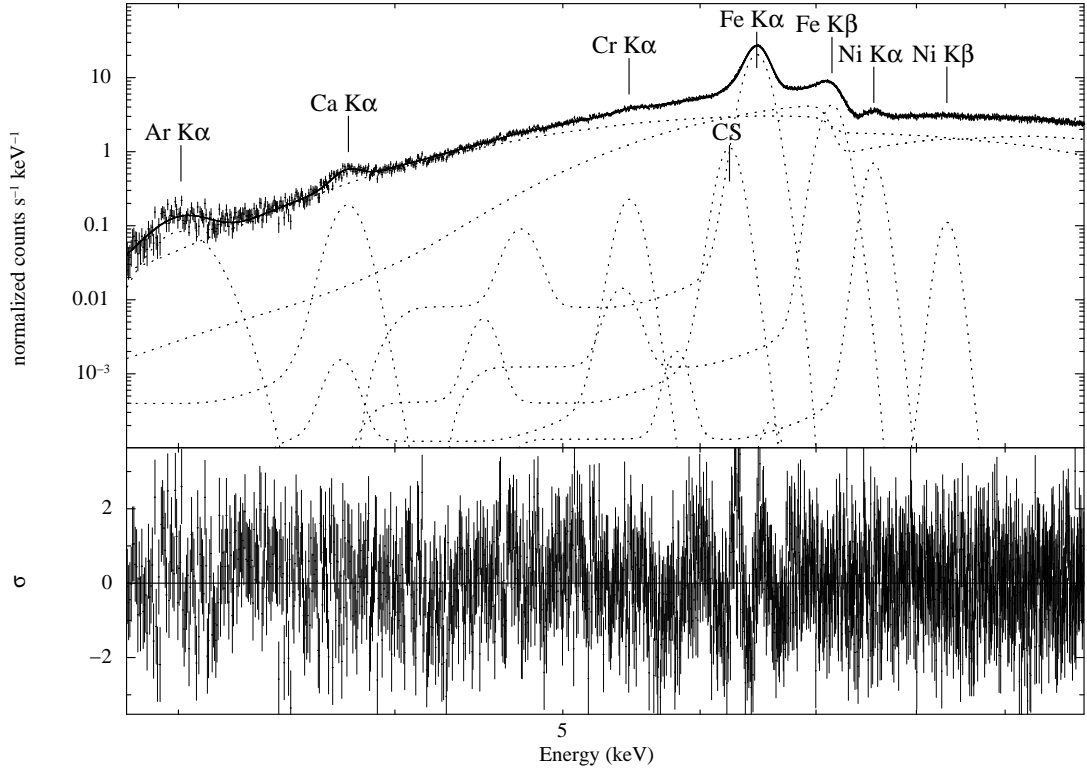}  
  \caption[]{\label{figure2} %
    \xmm with ID 0555200401 (2009) orbital phase averaged spectrum of \gx in the (2.8--10.0) keV band. \emph{Top panel}: data and best-fit model described by equation~(\ref{ec1}). \emph{Bottom panel} shows the residuals between the spectrum and the model. Spectral parameters can be found in Table~\ref{tab:4} (fourth column).
}
\end{figure}

The best-fit parameters for overall spectra of \gx and the corresponding uncertainties are summarized in Table~\ref{tab:4} where it is also included the equivalent width (EW) of the Gaussian emission lines. Figures~\ref{figure1} and \ref{figure2} show the data, the best-fit model described by equation~(\ref{ec1}), and the corresponding residuals.

The Gaussian emission lines were located at Ar K$\alpha$, Fe K$\alpha$, Fe K$\beta$, and Ni K$\alpha$, fluorescent line energies in the two \xmm observations, meanwhile the Gaussian emission lines located at Ca K$\alpha$, Cr K$\alpha$, and Ni K$\beta$ fluorescent line energies were only detected in the \xmm observation from 2009. Moreover, the Fe K$\alpha$ showed residuals when fitted with a single Gaussian and an additional Compton Shoulder was also included \citep{2003ApJ...597L..37W}. The \xmm observation from 2009 was also analysed by \citet{2011A&A...535A...9F}. They used different continuum models to fit this spectrum but they also added the same Gaussian emission lines used in this work. The \xmm observation from 2008 was also studied by \citet{2024MNRAS.527.2652R}. To describe this spectrum, these authors fitted a model as previously used by \citet{2011A&A...535A...9F} and identified the same K$\alpha$ emission lines, included the Compton Shoulder, and Fe K$\beta$ but Ni K$\beta$ was not detected.

The size of emitting region on the NS surface was obtained from the \emph{bbody} component by the equation:

\begin{equation}
R_\mathrm{bb}(\mathrm{km})=3.04 \times 10^{4} \frac{D\sqrt{F_\mathrm{bb}}}{T^{2}_\mathrm{bb}}\, ,
\label{ec2}
\end{equation}
where $D$ is the distance to the source in kpc, $F_\mathrm{bb}$ is the unabsorbed flux in erg s$^{-1}$ cm$^{-2}$ in the energy range [2.8--10.0] keV and $T_\mathrm{bb}$ is the temperature in keV. Taking into account the distance to the source given by \emph{GEDR3} $d (kpc) = 3.55^{+0.18}_{-0.16}$ it has been found that the radius of the emission region is consistent with a hot spot on the NS surface: from the 2008 observation, the calculated flux was $F_\mathrm{bb} = \left(1.40^{+0.17}_{-0.20}\right) \times 10^{-9}$ \ergscm and gave a radius $R_\mathrm{bb} = 0.41^{+0.10}_{-0.08}$ km, and from the 2009 observation, $F_\mathrm{bb} = \left(8.1^{+1.6}_{-0.9}\right) \times 10^{-10}$ \ergscm and $R_\mathrm{bb} = 0.27^{+0.07}_{-0.06}$ km, respectively.

The intrinsic bolometric X-ray luminosity can be expressed through the unabsorbed X-ray flux of the source as:
\begin{equation}
L_\mathrm{X}=4\, \pi\, D^{2}\, F_\mathrm{no\_abs}\, ,
\label{ec3}
\end{equation}
where $L_\mathrm{X}$ is the X-ray luminosity, $D$ is the distance to the source and $F_\mathrm{no\_abs}$ is the unabsorbed flux in the [2.8--10.0] keV energy band. The observed $F_\mathrm{no\_abs}$ for the 2008 and 2009 observations were $\left(6.9^{+0.8}_{-1.0}\right) \times 10^{-9}$ \ergscm and $\left(5.3^{+1.0}_{-0.6}\right) \times 10^{-9}$ \ergscm, respectively, corresponding to X-ray luminosities of $L_\mathrm{X_{2008}} = \left(1.1\pm0.3\right)\times10^{37}$ \ergs and $L_\mathrm{X_{2009}} = \left(8.1^{+2.4}_{-1.6}\right)\times10^{36}$ \ergs.

In Table~\ref{tab:4}, \emph{norm} represents the blackbody normalisation and is related to the luminosity of the X-ray source and the distance $\left(L_{39}/D^{2}_{10}\right)$. Matching this expression with the values of luminosities and distance to the source, the results were $0.08\pm0.10$ and $0.06^{+0.08}_{-0.07}$ from 2008 and 2009 observations, respectively, which is consistent with the fit values taking the uncertainties into account.

\begin{table}[h!tbp]\centering
{\scriptsize
  \setlength{\tabnotewidth}{0.5\columnwidth}
  \tablecols{3}
  % Stretch the space between table columns 
  \setlength{\tabcolsep}{2.8\tabcolsep}
  \caption{Best-fit model parameters for the averaged spectra\tabnotemark{a}} \label{tab:4}
 \begin{tabular}{lrrr}
    \toprule
    Component & \multicolumn{1}{c}{Parameter} & \multicolumn{1}{c}{Value (ID 0555200301)} & \multicolumn{1}{c}{Value (ID 0555200401)} \\
    \midrule
\emph{tbnew} & $N^{1}_{H}$ [10$^{22}$ atoms cm$^{-2}$] & 191$^{+19}_{-9}$ & 183$^{+12}_{-10}$ \\
 & $N^{2}_{H}$ [10$^{22}$ atoms cm$^{-2}$] & 42.2$^{+1.1}_{-1.0}$ & 53.3$^{+2.2}_{-1.7}$ \\ \midrule
\emph{Power law} & Photon index $\Gamma$ & 1.2$\pm$0.3 &1.03$^{+0.13}_{-0.12}$ \\
 & norm [keV$^{-1}$ s$^{-1}$ cm$^{-2}$] & 0.7 $^{+0.7}_{-0.3}$ & 0.40 $^{+0.14}_{-0.10}$ \\ \midrule
\emph{bbody} & $kT$ [keV] & 3.14$^{+0.21}_{-0.11}$ & 3.37$\pm$0.22 \\
 & norm  [$L_{39}/D^{2}_{10}$] & 0.041$^{+0.005}_{-0.006}$ & 0.026$^{+0.005}_{-0.002}$\\ \midrule
Ar K$\alpha$ & Line E [keV]& 2.96$\pm$0.04 & 3.013$\pm$0.012 \\
 & $\sigma$ [keV] &  0.005$^{+0.010}_{-0.005}$ & 0.092$^{+0.021}_{-0.018}$ \\
 &  EW [keV]& 0.11$^{+0.06}_{-0.03}$ & 0.82$^{+0.09}_{-0.07}$ \\
 & norm [10$^{-5}$ ph s$^{-1}$ cm$^{-2}$]& 2.8$^{+1.5}_{-0.7}$ & 4.7$^{+0.5}_{-0.4}$ \\ \midrule
Ca K$\alpha$ & Line E [keV]& & 3.756$\pm$0.008 \\
 & $\sigma$ [keV] & & 0.021$^{+0.018}_{-0.021}$ \\
 & EW [keV]& & 0.094$\pm$0.009 \\
 & norm [10$^{-5}$ ph s$^{-1}$ cm$^{-2}$]& & 6.4$^{+0.6}_{-0.6}$ \\ \midrule
Cr K$\alpha$ & Line E [keV]& & 5.463$^{+0.021}_{-0.015}$ \\
 & $\sigma$ [keV] & & 0.01 (frozen) \\
 & EW [keV]& & 0.0117$^{+0.0023}_{-0.0022}$ \\
 & norm [10$^{-5}$ ph s$^{-1}$ cm$^{-2}$]& & 8.1$^{+1.6}_{-1.5}$ \\ \midrule
Compton Shoulder & Line E [keV]& 6.24 (frozen) & 6.24 (frozen)\\
 & $\sigma$ [keV] & 0.01 (frozen) & 0.01 (frozen) \\
 & EW [keV]& 0.0207$^{+0.0024}_{-0.0021}$ &  0.0163$^{+0.0011}_{-0.0007}$ \\
 & norm [10$^{-4}$ ph s$^{-1}$ cm$^{-2}$]& 7.7$^{+0.9}_{-0.8}$ & 4.5$^{+0.3}_{-0.2}$ \\ \midrule
Fe K$\alpha$ & Line E [keV]& 6.4388$^{+0.0017}_{-0.0009}$ & 6.4698$^{+0.0007}_{-0.0004}$ \\
 & $\sigma$ [keV]& 0.0462$^{+0.0019}_{-0.0022}$ & 0.0503$^{+0.0008}_{-0.0009}$\\
 & EW [keV]& 0.528$^{+0.005}_{-0.006}$ & 0.701$\pm$0.003 \\
 & norm [10$^{-2}$ ph s$^{-1}$ cm$^{-2}$]& 1.159$^{+0.012}_{-0.014}$ & 1.016$\pm$0.005 \\ \midrule
Fe K$\beta$ & Line E [keV]& 7.121$^{+0.005}_{-0.004}$ & 7.1518$^{+0.0019}_{-0.0012}$ \\
 & $\sigma$ [keV]& 0.063$^{+0.006}_{-0.007}$ & 0.056$^{+0.002}_{-0.003}$ \\
 & EW [keV]& 0.134$\pm$0.005 & 0.223$\pm$0.003 \\
 & norm [10$^{-3}$ ph s$^{-1}$ cm$^{-2}$]& 2.49$^{+0.09}_{-0.10}$ & 2.64$^{+0.03}_{-0.04}$ \\ \midrule
Ni K$\alpha$ & Line E [keV]& 7.513$^{+0.015}_{-0.018}$ & 7.558$^{+0.003}_{-0.012}$ \\
 & $\sigma$ [keV]& 0.03$\pm$0.03 & 0.00$^{+0.03}_{-0.00}$ \\
 & EW [keV]& 0.027$\pm$0.005 & 0.0422$\pm$ 0.0023 \\
 & norm [10$^{-4}$ ph s$^{-1}$ cm$^{-2}$]& 4.0$^{+0.8}_{-0.7}$ & 4.03$\pm$0.22 \\ \midrule
Ni K$\beta$ & Line E [keV]& & 8.33$\pm$0.04 \\
 & $\sigma$ [keV] & & 0.03$^{+0.07}_{-0.03}$ \\
 & EW [keV]& & 0.008$\pm$0.003 \\
 & norm [10$^{-5}$ ph s$^{-1}$ cm$^{-2}$]& & 9$\pm$3 \\ \midrule
$\chi^{2}_{r}$ & & \multicolumn{1}{c}{$\chi^{2}$/(d.o.f.) = 1531/1420 = 1.1}  & \multicolumn{1}{c}{$\chi^{2}$/(d.o.f.) = 1711/1412 = 1.2}  \\
    \bottomrule
    \tabnotetext{a}{Parameters for equation~(\ref{ec1}). Observations ID 0555200301 (third column) and 0555200401 (fourth column). EW represents the equivalent width of the emission line. Uncertainties are given at the 90\% confidence limit and d.o.f is degrees of freedom.}
  \end{tabular}
  }
\end{table}

An intense Fe K$\alpha$ emission line (EW $\approx$ [0.5, 0.7] keV) with a global hydrogen column above 10$^{24}$ cm$^{-2}$ has been found in the analysis and it is compatible with \citet{1988PASJ...40..197L,1989MNRAS.236..603L}, %(Leahy et al., 1988, 1989) 
and \citet{2011A&A...535A...9F} who found in this X-ray region a strong emission line Fe K$\alpha$ visible together with a hydrogen column above 10$^{24}$ cm$^{-2}$. 

The flux ratio between the Fe K$\alpha$ and Fe K$\beta$ allowed us to derive that the ionisation state of iron varied from Fe \textsc{xii-xiii} $\left(\right.\mathrm{Fe}$ K$\beta$/Fe K$\alpha = \left.0.215^{+0.010}_{-0.011}\right)$ in 2008 observation to Fe $>$ \textsc{xiv} (Fe K$\beta$/Fe K$\alpha = 0.260^{+0.004}_{-0.005}$) in 2009 observation. These results pointed out that only mildly ionised iron was present in the plasma \citep{2011A&A...535A...9F}. On the other hand, the flux ratio between Ni K$\alpha$ and Ni K$\beta$ (Ni K$\beta$/Ni K$\alpha = 0.23\pm0.09$) was slightly consistent with measurements in solid state metals within errors \citep{2009PhRvA..80e2503H} in 2009 observation.

The results mentioned above show a slightly different values between parameters from one observation to the other. In order to further study these differences, the pre-periastron flare has been analysed in the long term to find how the spectral parameters evolve with time. In addition, \gx also shows a periodic near-apastron flare with lower intensity than the pre-periastron flare and it is also included in this long term analysis. For such studies, it is also very convenient to have spectra measured at different orbital phases with uniform phase coverage. The characteristics of the \emph{Monitor of All Sky X-ray Image} (\emph{MAXI}), moderate energy resolution and all-sky coverage, are well suited for elaborate studies of the orbital phase resolved spectra of bright X-ray sources. The process which was used to obtain the data is explained in \S~\ref{timing}.

Observations, data and timing analysis are described in \S~\ref{timing}, \maxi spectral analysis for the pre-periastron and near-apastron flares are  discussed in \S~\ref{spectra}, and \S~\ref{conclusion} contains the summary of the main results.

\begin{table}[h!tb]%[scale=2.0,width=2\columnwidth]
 \caption{Ephemeris data used for \maxi's timing calculations}
    \centering
    \begin{tabular}{lccc} \toprule\toprule     
$T_{\mathrm{0,periastron\_passage}}$ (MJD) & $48802.79\pm 0.12$ \\
$P_{\mathrm{orb}}$ (d)  & $41.498 \pm 0.002$ \citep{1997ApJ...479..933K} \\

     \bottomrule
    \end{tabular}
    
    \label{tab:1}
\end{table}

\section{Timing analysis}
\label{timing}

\begin{figure}[hbtp]
  \centering
  \includegraphics[angle=-90.0,width=\columnwidth]{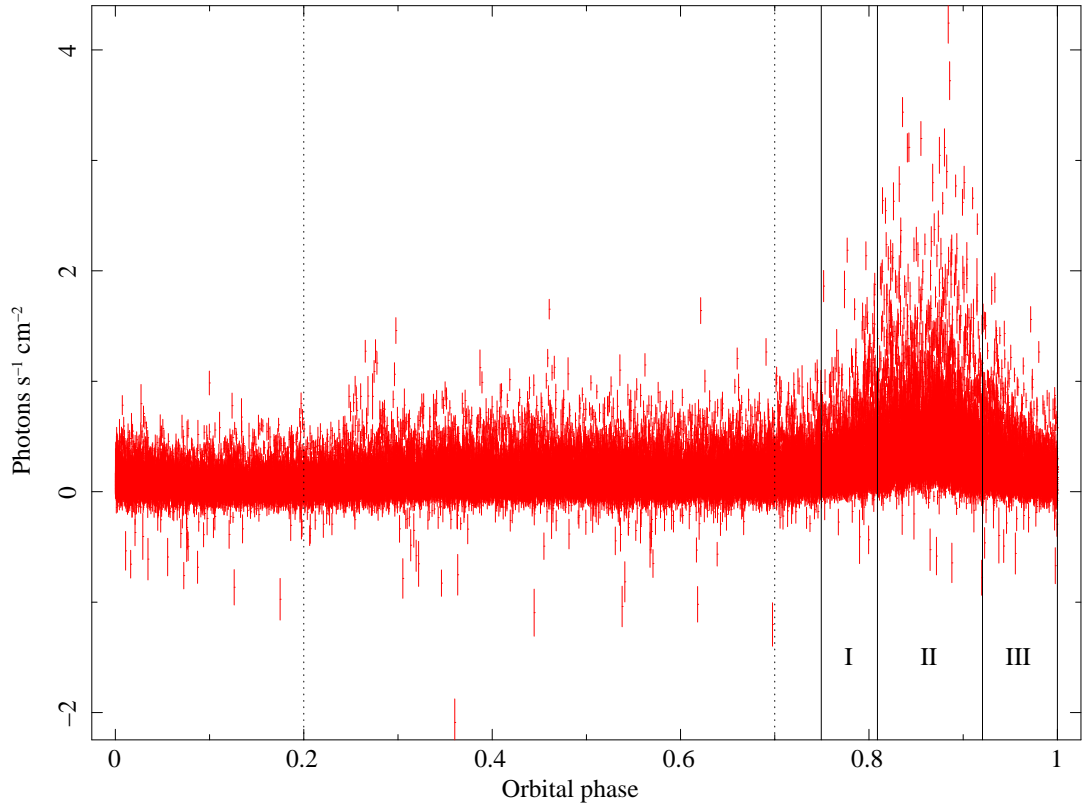}  
  \caption[]{\label{figure3} %
    Folded and background subtracted light curve in the (2.0--20.0) keV energy range. The light curve was folded over the orbital period and binned into $\sim$ 450 phase bins. The selection criteria for the apastron flare from MAXI data were:
orbital phase range = [0.20--0.70] and (2-20) keV flux $>$ 0.5 photons cm$^{-2}$ s$^{-1}$.  
}
\end{figure}

The orbital phase of the pre-periastron flare has been divided into three sections to perform the spectral analysis. To define them the \maxi (2.0--20.0) keV light curve with a time bin of 0.02 days\footnote{http://maxi.riken.jp/pubdata/v7lrkn/J1226-627/index.html} has been folded using the ephemeris from \citet{1997ApJ...479..933K}, %Koh et al.
which are reported in Table~\ref{tab:1}, and a circular orbit has been assumed because we wanted to have an approximation of the results. Figure~\ref{figure3} shows the light curve folded over the \gx orbital period where three sections have been identified by vertical lines: I pre-flare, II flare and III post-flare. As a consequence of the low intensity level, only one section has been used to define the near-apastron flare at [0.20--0.70] orbital phase range. For each orbit, the Modified Julian Dates (MJDs) for every orbital phase range of the pre-periastron flare have been calculated. For each section, data were accumulated over ten consecutive orbits to extract spectra with acceptable signal-to-noise ratio. As an example, Table~\ref{tab:2} summarizes MJDs of the first ten orbits for pre-flaring, flaring and post-flaring which were used as good time intervals (GTIs) to extract the spectra. One possible implication of the circular orbit assumption is that the MJDs would be affected by the approach of the orbit and there could be inaccuracies that could affect the conclusions obtained. However, the spectra are consistent from one set to the next one (see Figures~\ref{figure4}-\ref{figure6}) and this indicates that the assumption of a  circular orbit is useful in the system \gx when analysing the pre-periastron flare.

\begin{figure}[htb]
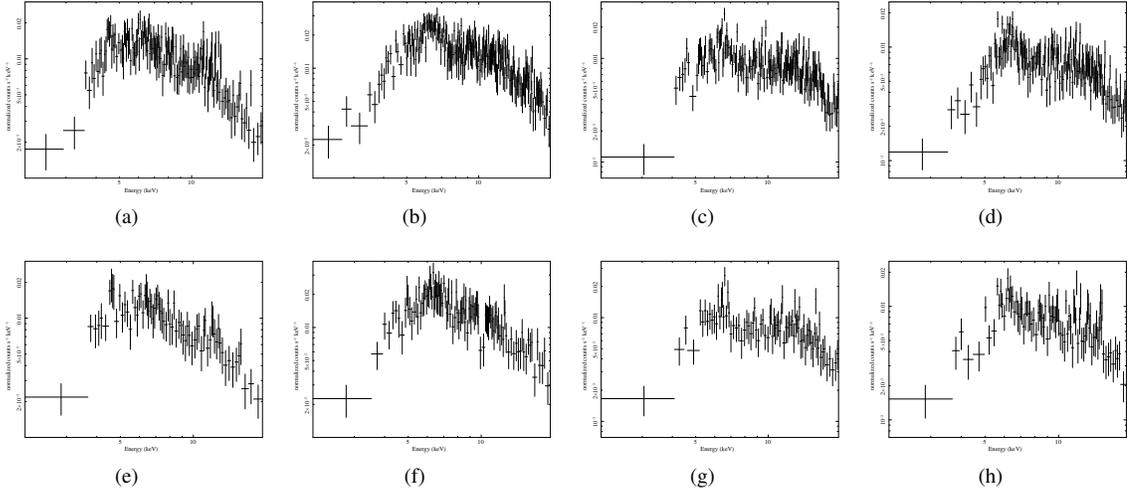

\centering
  \subfloat[]{%
    \includegraphics[angle=-90.0,width=.24\textwidth]{terceras_orbitas_inicio.ps}}\hfill
  \subfloat[]{%
    \includegraphics[angle=-90.0,width=.24\textwidth]{novenas_orbitas_inicio.ps}}\hfill
  \subfloat[]{%
    \includegraphics[angle=-90.0,width=.24\textwidth]{quintas_orbitas_final.ps}}\hfill
  \subfloat[]{%
    \includegraphics[angle=-90.0,width=.24\textwidth]{octavas_orbitas_final.ps}}\\
  \subfloat[]{%
    \includegraphics[angle=-90.0,width=.24\textwidth]{terceras_orbitas_inicio_mitad.ps}}\hfill
  \subfloat[]{%
    \includegraphics[angle=-90.0,width=.24\textwidth]{novenas_orbitas_inicio_mitad.ps}}\hfill
  \subfloat[]{%
    \includegraphics[angle=-90.0,width=.24\textwidth]{quintas_orbitas_final_mitad.ps}}\hfill
  \subfloat[]{%
    \includegraphics[angle=-90.0,width=.24\textwidth]{octavas_orbitas_final_mitad.ps}}\hfill
  \caption{MAXI/GSC spectra of \gx in the (2.0--20.0) keV band. (a) and (e): pre-flare, third extraction. (b) and (f) pre-flare, ninth extraction. (c) and (g): post-flare, fifth extraction. (d) and (h): post-flare, eighth extraction. Spectra (a)$-$(d) correspond to the accumulation of ten orbits. Spectra (e)$-$(h) correspond to the accumulation of five orbits. Units of the x-axis: Energy (keV). Units of the y-axis: normalized counts s$^{-1}$ keV$^{-1}$.}\label{orbitas}
\end{figure}

Figure~\ref{orbitas} shows a comparison between spectra obtained by accumulating ten orbits (top spectra) and spectra obtained by accumulating five orbits (bottom spectra).

\begin{table}
 \caption{First ten orbits: pre-flare, flare and post-flare} 
  \centering
  \begin{tabular}{c c c c c c} \toprule
 Pre-flare ($\sim$[0.7493, 0.8090]) & Flare  ($\sim$[0.8090, 0.9205]) & Post-flare ($\sim$[0.9205, 0.9984])\\ \midrule
 $MJD_{i}-MJD_{f}$ & $MJD_{i}-MJD_{f}$ & $MJD_{i}-MJD_{f}$ \\ \midrule
 55141.582121$-$55144.056111 & 55144.056111$-$55148.686661 & 55148.686661$-$55151.921117 \\
 55183.080121$-$55185.554111 & 55185.554111$-$55190.184661 & 55190.184661$-$55193.419117 \\
55224.578121$-$55227.052111 & 55227.052111$-$55231.682661 & 55231.682661$-$55234.917117 \\
55266.076121$-$55268.550111 & 55268.550111$-$55273.180661 & 55273.180661$-$55276.415117 \\
55307.574121$-$55310.048111 & 55310.048111$-$55314.678661 & 55314.678661$-$55317.913117 \\
55349.072121$-$55351.546111 & 55351.546111$-$55356.176661 & 55356.176661$-$55359.411117 \\
55390.570121$-$55393.044111 & 55393.044111$-$55397.674661 & 55397.674661$-$55400.909117 \\
55432.068121$-$55434.542111 & 55434.542111$-$55439.172661 & 55439.172661$-$55442.407117 \\
55473.566121$-$55476.040111 & 55476.040111$-$55480.670661 & 55480.670661$-$55483.905117 \\
55515.064121$-$55517.538111 & 55517.538111$-$55522.168661 & 55522.168661$-$55525.403117 \\
    \bottomrule
     
  \end{tabular}
  \label{tab:2}
\end{table}

\begin{figure}[hbtp]
  \centering
  \includegraphics[angle=0.0,width=\columnwidth]{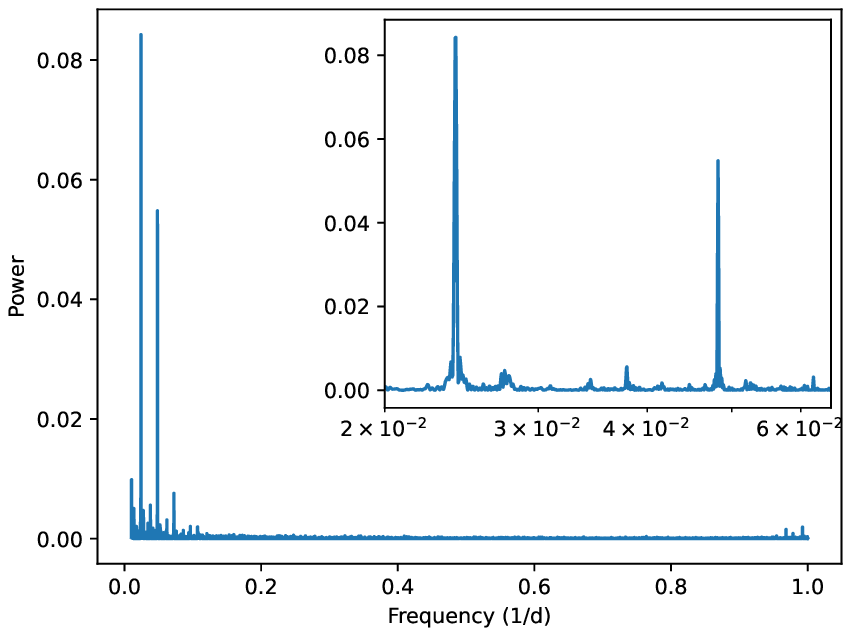}  
  \caption[]{\label{periodogram} %
     Lomb-Scargle periodogram of the (4.0--10.0) keV light curve. The plot on the right top corner has been added to show only lower frequencies where the peaks lie.
  } 
\end{figure}

In the context of \maxi data, the passbands often used are called ``soft'' (2.0--4.0) keV, ``medium'' (4.0--10.0) keV, and ``hard'' (10--20.0) keV bands. To search for pre-periastron and near-apastron changes in the X-ray emission, light curves in the (2.0--4.0) keV, (4.0--10.0) keV, (10--20.0) keV and (5.7--7.5) keV (iron complex emission lines) energy bands were extracted and analysed using \emph{Astropy}, a collection of software packages which are included in \emph{Python} \citep{astropy:2022}. In each of these light curves a \emph{Lomb-Scargle} periodogram \citep{1989ApJ...338..277P} was applied and the error in the period is approximately the area where the peak is at the 90\% of its value. It was apparent that the light curves were similar to each other and, in this work, only the periodogram of the light curve (4--10) keV is shown (see Figure~\ref{periodogram}). As can be seen from Figure~\ref{periodogram}, the highest peak (power $\approx$ 0.084) is at (41.4$\pm$0.5) days which corresponds to the rotational period of the binary system determined by \citet{1997ApJ...479..933K}. The second peak (power $\approx$ 0.055) in the power density spectrum at $\approx$20.7 days is potentially a harmonic of the orbital period.

\begin{figure}[hbtp]
  \centering
  \includegraphics[angle=-90.0,width=\columnwidth]{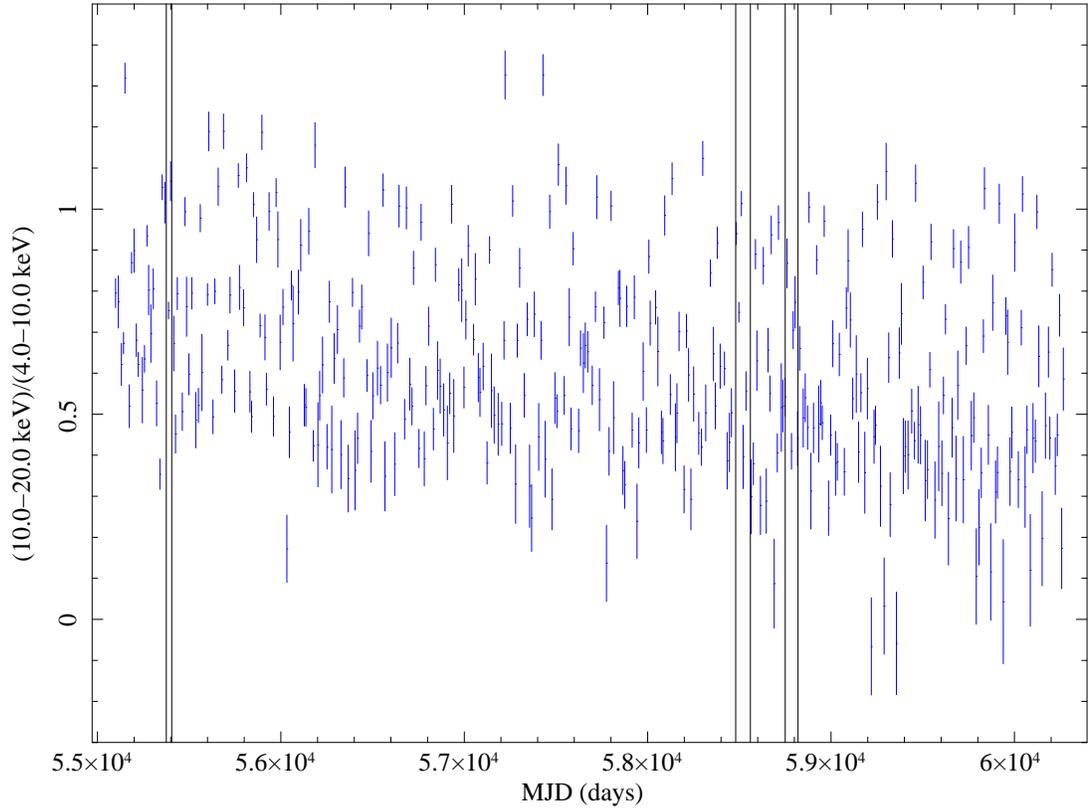}  
  \caption[]{\label{hardness_ratio_v2} %
   Hardness ratio $hard/medium = (10.0-20.0\, \, \mathrm{keV})/(4.0-10.0\, \, \mathrm{keV})$ using weighted average. The spin-up episodes are indicated by two vertical lines: (a) from MJD 55375 to 55405, (b) from MJD 58480 to 58560 and (c) from MJD 58750 to 58820.
  } 
\end{figure}

\begin{figure}[hbtp]
  \centering
  \includegraphics[angle=-90.0,width=\columnwidth]{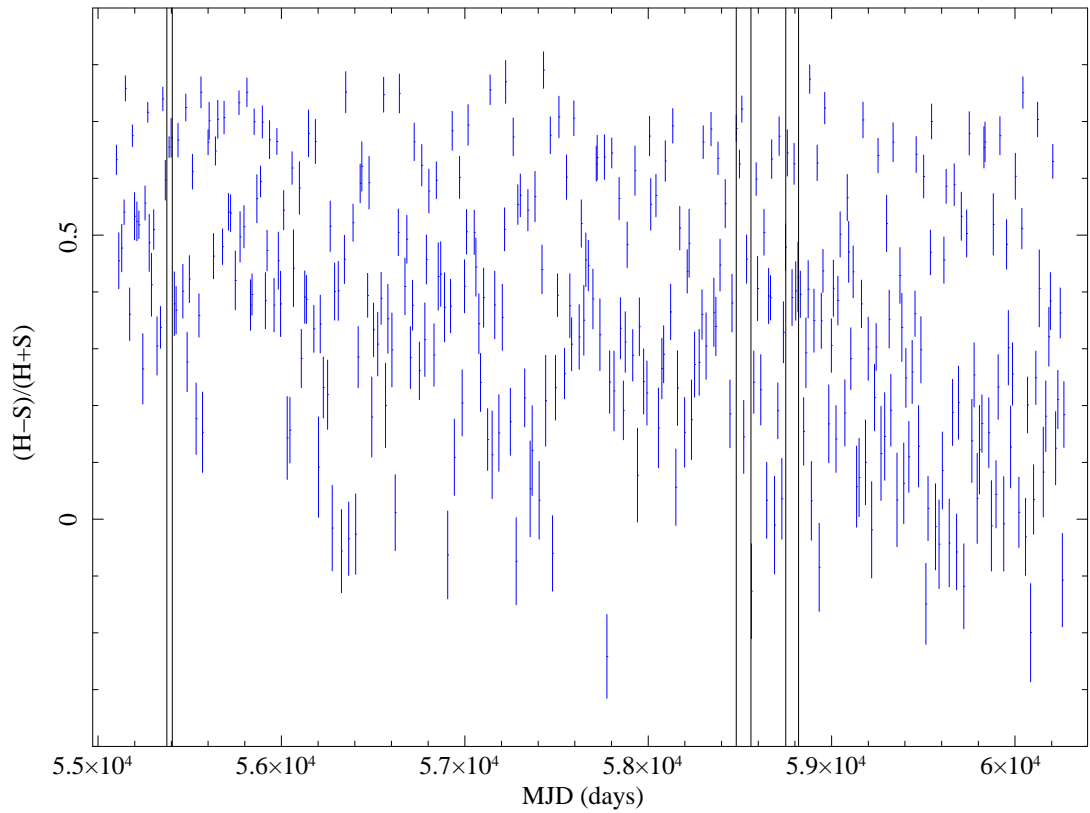}  
  \caption[]{\label{hardness_curves_v2} %
    Hardness curve $(\mathrm{medium}-\mathrm{soft})/(\mathrm{medium}+\mathrm{soft})$ using weighted average. The spin-up episodes are indicated by two vertical lines: (a) from MJD 55375 to 55405, (b) from MJD 58480 to 58560 and (c) from MJD 58750 to 58820.
  } 
\end{figure}

The hardness ratio is a specially useful tool to quantify and characterise the source spectrum. Therefore, simple hardness ratio (hard/medium) and fractional difference hardness ratio (medium$-$soft)/(medium$+$soft) have been calculated using the weighted average over 150 bins and plotted in Figures~\ref{hardness_ratio_v2} and \ref{hardness_curves_v2}, respectively. The weighting factor here for computing the average HR is the error weighted average, where the errors have been calculated using the general method of getting formulas for propagating errors. Both graphs show that the ratios continues to oscillate around the average, with no clear trend.

\begin{figure}[hbtp]
  \centering
  \includegraphics[angle=-90.0,width=\columnwidth]{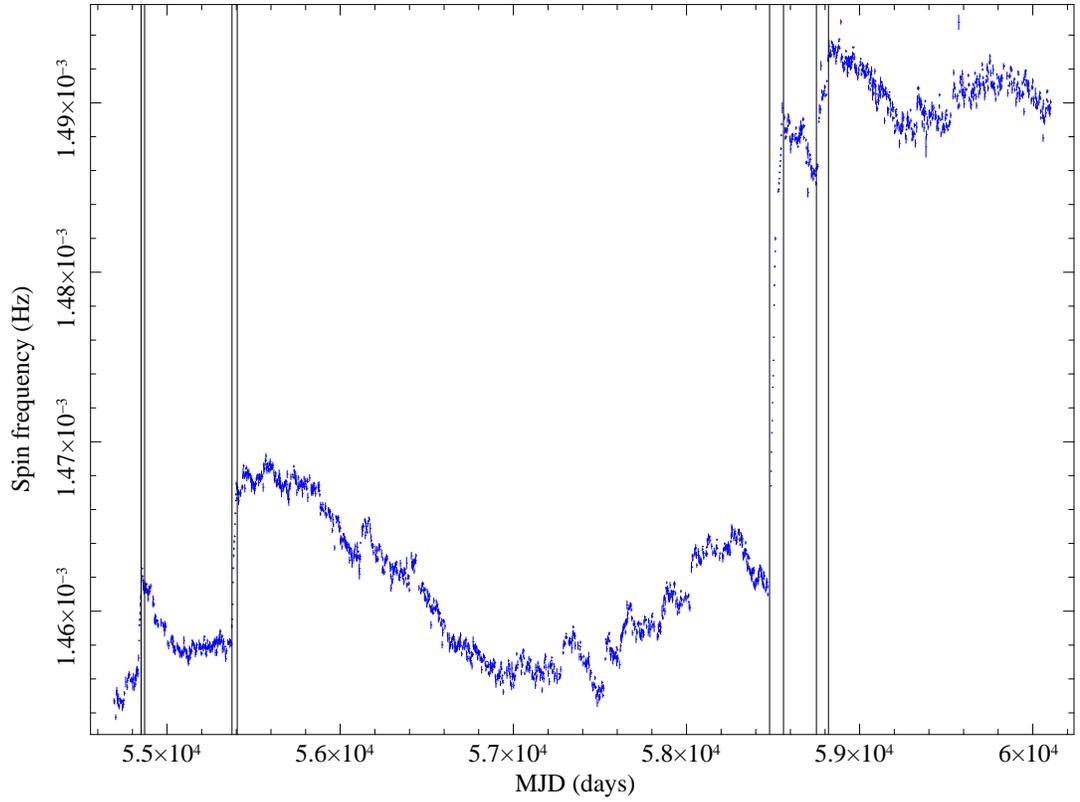}  
  \caption[]{\label{spin_frequency_3} %
    Spin frequency versus the MJD for the system \gx using the data from \fermi. There are four spin-up episodes which are indicated by two vertical lines: (a) from MJD 54850 to 54870, (b) from MJD 55375 to 55405, (c) from MJD 58480 to 58560 and (d) from MJD 58750 to 58820.
  } 
\end{figure}

Figure~\ref{spin_frequency_3} shows the pulse frequency history of \gx observed with \fermi Accreting Pulsars Program\footnote{http://gammaray.nsstc.nasa.gov/gbm/science/pulsars} \citep{2020ApJ...896...90M}. Every two vertical lines on this plot represents a spin-up episode in the history of observations of the source which have been associated with the formation of a transient accretion disc \citep{1997ApJ...479..933K,2019A&A...629A.101N}. Two of these irregular spin-up episodes can be related to the apastron passages (see \S~\ref{apastron_flare}).

\section{Spectral analysis}
\label{spectra}

\subsection{Pre-periastron flare spectra}
\label{pre_periastron_flare}

\begin{figure}[htb]
\centering
  \subfloat[]{%
    \includegraphics[angle=-90.0,width=.24\textwidth]{primeras_orbitas_inicio.ps}}\hfill
  \subfloat[]{%
    \includegraphics[angle=-90.0,width=.24\textwidth]{segundas_orbitas_inicio.ps}}\hfill
  \subfloat[]{%
    \includegraphics[angle=-90.0,width=.24\textwidth]{terceras_orbitas_inicio_fit.ps}}\hfill
  \subfloat[]{%
    \includegraphics[angle=-90.0,width=.24\textwidth]{cuartas_orbitas_inicio.ps}}\\
  \subfloat[]{%
    \includegraphics[angle=-90.0,width=.24\textwidth]{quintas_orbitas_inicio.ps}}\hfill
  \subfloat[]{%
    \includegraphics[angle=-90.0,width=.24\textwidth]{sextas_orbitas_inicio.ps}}\hfill
  \subfloat[]{%
    \includegraphics[angle=-90.0,width=.24\textwidth]{septimas_orbitas_inicio.ps}}\hfill
  \subfloat[]{%
    \includegraphics[angle=-90.0,width=.24\textwidth]{octavas_orbitas_inicio.ps}}\\
  \subfloat[]{%
    \includegraphics[angle=-90.0,width=.24\textwidth]{novenas_orbitas_inicio_fit.ps}}\hfill
  \subfloat[]{%
    \includegraphics[angle=-90.0,width=.24\textwidth]{decimas_orbitas_inicio.ps}}\hfill
  \caption{\maxi spectra of \gx in the (2.0--20.0) keV band corresponding to the pre-flare ($\sim$[0.7493, 0.8090]), model defined by equation ~(\ref{ec6}). The same orbital phase range was accumulated in groups of ten consecutive orbits to obtain each spectrum. Units of the x-axis: Energy (keV). Units of the y-axis: normalized counts s$^{-1}$ keV$^{-1}$.}\label{figure4}
\end{figure}

\begin{figure}[htb]
\centering
  \subfloat[]{%
    \includegraphics[angle=-90.0,width=.24\textwidth]{primeras_orbitas_fase_intermedia.ps}}\hfill
  \subfloat[]{%
    \includegraphics[angle=-90.0,width=.24\textwidth]{segundas_orbitas_fase_intermedia.ps}}\hfill
  \subfloat[]{%
    \includegraphics[angle=-90.0,width=.24\textwidth]{terceras_orbitas_fase_intermedia.ps}}\hfill
  \subfloat[]{%
    \includegraphics[angle=-90.0,width=.24\textwidth]{cuartas_orbitas_fase_intermedia.ps}}\\
  \subfloat[]{%
    \includegraphics[angle=-90.0,width=.24\textwidth]{quintas_orbitas_fase_intermedia.ps}}\hfill
  \subfloat[]{%
    \includegraphics[angle=-90.0,width=.24\textwidth]{sextas_orbitas_fase_intermedia.ps}}\hfill
  \subfloat[]{%
    \includegraphics[angle=-90.0,width=.24\textwidth]{septimas_orbitas_fase_intermedia.ps}}\hfill
  \subfloat[]{%
    \includegraphics[angle=-90.0,width=.24\textwidth]{octavas_orbitas_fase_intermedia.ps}}\\
  \subfloat[]{%
    \includegraphics[angle=-90.0,width=.24\textwidth]{novenas_orbitas_fase_intermedia.ps}}\hfill
  \subfloat[]{%
    \includegraphics[angle=-90.0,width=.24\textwidth]{decimas_orbitas_fase_intermedia.ps}}\hfill
  \caption{\maxi spectra of \gx in the (2.0--20.0) keV band corresponding to the flare ($\sim$[0.8090, 0.9205]), model defined by equation ~(\ref{ec6}). The same orbital phase range was accumulated in groups of ten consecutive orbits to obtain each spectrum. Units of the x-axis: Energy (keV). Units of the y-axis: normalized counts s$^{-1}$ keV$^{-1}$.
}\label{figure5}
\end{figure}

\begin{figure}[htb]
\centering
  \subfloat[]{%
    \includegraphics[angle=-90.0,width=.24\textwidth]{primeras_orbitas_final.ps}}\hfill
  \subfloat[]{%
    \includegraphics[angle=-90.0,width=.24\textwidth]{segundas_orbitas_final.ps}}\hfill
  \subfloat[]{%
    \includegraphics[angle=-90.0,width=.24\textwidth]{terceras_orbitas_final.ps}}\hfill
  \subfloat[]{%
    \includegraphics[angle=-90.0,width=.24\textwidth]{cuartas_orbitas_final.ps}}\\
  \subfloat[]{%
    \includegraphics[angle=-90.0,width=.24\textwidth]{quintas_orbitas_final_fit.ps}}\hfill
  \subfloat[]{%
    \includegraphics[angle=-90.0,width=.24\textwidth]{sextas_orbitas_final.ps}}\hfill
  \subfloat[]{%
    \includegraphics[angle=-90.0,width=.24\textwidth]{septimas_orbitas_final.ps}}\hfill
  \subfloat[]{%
    \includegraphics[angle=-90.0,width=.24\textwidth]{octavas_orbitas_final_fit.ps}}\\
  \subfloat[]{%
    \includegraphics[angle=-90.0,width=.24\textwidth]{novenas_orbitas_final.ps}}\hfill
  \subfloat[]{%
    \includegraphics[angle=-90.0,width=.24\textwidth]{decimas_orbitas_final.ps}}\hfill
  \caption{\maxi spectra of \gx in the (2.0--20.0) keV band corresponding to the post-flare ($\sim$[0.9205, 0.9984]), model defined by equation ~(\ref{ec6}). The same orbital phase range was accumulated in groups of ten consecutive orbits to obtain each spectrum. Units of the x-axis: Energy (keV). Units of the y-axis: normalized counts s$^{-1}$ keV$^{-1}$.
}\label{figure6}
\end{figure}

The spectra were obtained using the \maxi on-demand process\footnote{http://maxi.riken.jp/mxondem}, all of them were analysed and modelled with the \textsc{XSPEC} \citep{1996ASPC..101...17A} package and the energy range used for spectral fitting was (2.0--20.0) keV. Both phenomenological and physical models commonly applied to accreting X-ray pulsars have been tested. Models with power law components have been developed to \hmxbs, such as \centau \citep{1996PASJ...48..425E} and \gx \citep[]{2011A&A...535A...9F,2021MNRAS.501.2522J}. First of all, traditional models like powerlaw with a high energy cut-off or with a Fermi-Dirac cut-off were explored. Although these models reasonably well reproduce the observed flare spectra between (2.0--20.0) keV ($\chi^{2}_{r} = [0.96-1.32]$), it cannot offer a good statistical and/or physical solution in all spectra because the parameters of the high energy cut-off and the photon index ($\Gamma = [-0.3, 0.6]$) could not be constrained well. As the principal aim is to describe with a consistent model all spectra these models were rejected. Then, it was described the orbital phase-averaged spectra using a simple model with two power laws described by equations~(\ref{ec4}) and ~(\ref{ec5}). A fast charged particle traversing a region containing a strong magnetic field will change direction because the magnetic field exerts a force perpendicular to the direction of motion. Because the velocity vector changes, the charged particle is accelerated and consequently emits electromagnetic energy which is called synchroton radiation. Observed polarisation in the radiation is usually a proof of synchrotron emission \citep{astronomia_2005}. The usual spectra form can be described by a power law $F(E)=B\, E^{-\Gamma}$, where $B$ represents a constant and $\Gamma$ is the photon index. The more positive the value of the $\Gamma$, the softer is the spectrum. The scattering of X-rays by interstellar dust softens the spectrum by $E^{-2}$ relative to the source spectrum and soft X-rays are scattered more strongly than hard X-rays. In the energy range (0.1--10) keV, the main interaction between X-rays and matter is the photoelectric effect whose cross-section varies with energy as $\sim Z^3\, E^{-3}$. Consequently, absorption is greatest at low energies and in high-\emph{Z} materials. Thus, the energy spectra have been fitted by two power-law components, where each photon index indicates the source spectrum and the scattering or absorption spectrum, respectively. Initially, the orbital phase-resolved spectra were fitted with a composite model of two absorbed power laws:

\begin{equation}
  F(E) = \mathit{tbnew} \times B\, E^{-\gamma} + \mathit{tbnew} \times B\, E^{-(\gamma + 2.0)} + \mathit{GL}\, ,
  \label{ec4}
\end{equation}

\begin{equation}
  F(E) = \mathit{tbnew} \times B E^{-\gamma} + \mathit{tbnew} \times B E^{-(\gamma + 3.0)} + \mathit{GL}\, .
  \label{ec5}
\end{equation}

Although both models described above fitted the averaged spectra between (2.0--20.0) keV with similar statistical confidence (equation~\ref{ec4}, $\chi^{2}_{r}$ = [0.86--1.10]; equation~\ref{ec5}, $\chi^{2}_{r}$ = [0.89--1.18]), none of them could offer a consistent astrophysical solution to all the resolved spectra. For example, in the flare spectra $N^{2}_{H}$ values were above $200 \times 10^{22}$ cm$^{-2}$ in eight spectra (equation~(\ref{ec4}), $N^{2}_{H} \sim \left(231-10^{6}\right)\times 10^{22}$ cm$^{-2}$; equation~(\ref{ec5}), $N^{2}_{H} \sim \left(266-10^{6}\right)\times 10^{22}$ cm$^{-2}$), which were not consistent with the $N_{H}$ values $\left(10-80\right) \times 10^{22}$ cm$^{-2}$ obtained in the \maxi analysis of \citet{2014MNRAS.441.2539I}.

During flare episodes it is expected that the surface temperature of the neutron star can reach several million degrees and, therefore, it will emit blackbody radiation with photons in the X-ray range. The blackbody component has been used to describe the soft energy band spectra of \hmxbs like \centau using data from \xmm \citep{2021MNRAS.501.5892S} and \maxi \citep{2022RMxAA..58..355T}. The overall \maxi spectra of the source was modelled with an absorbed \emph{bbody} component. The spectral shape showed evidence for Fe K-shell absorption edge at $\approx$ 7.1 keV \citep{1996ApJ...463..726S,2002ApJ...574..879E,2021MNRAS.501.2522J}, therefore, an edge component fixed at this energy was added. The following model was applied to describe all observational data:

\begin{equation}
  F(E) = \mathit{tbnew} \times \mathit{bbody} \times \mathit{edge} + \mathit{GL}\, ,
  \label{ec6}
\end{equation}

where the Gaussian line was also included to account for the iron fluorescent emission line at $\sim$6.4 keV, if present. The model gave a good statistical description (pre-flare, $\chi^{2}_{r}=0.91-1.22$; flare, $\chi^{2}_{r}=0.89-1.07$; post-flare, $\chi^{2}_{r}=0.85-1.18$). In Figures~\ref{figure4}-\ref{figure6} the ten \maxi spectra for each pre-periastron section, the best-fitting model, and residuals to the best-fitting model are shown.

Some spectra showed a low-energy excess below $\sim$4.0 keV as it can see in Figure~\ref{figure4}, images (b), (c), (g), (i) and (j); Figure~\ref{figure5}, image (g); and Figure~\ref{figure6}, images (g), (h) and (j). This soft excess could be produced by a transitory structure in the line of sight because it was not a permanent effect in all X-ray spectra. Therefore, a possible explanation may be the presence of a transitory disc which enhances the accretion of matter. X-ray pulsars such as \gx tend to be spinning up because the accreted material from the optical star has an angular momentum which is eventually transferred to the compact object. The higher the X-ray luminosity, the more material is accreted by the pulsar, and therefore the faster it will spin. The strongest spin-up event of \gx so far (between 2018 December and 2019 March) pointed out to an accretion due to both direct accretion from the stellar wind and a temporary accretion disc \citep{2019A&A...629A.101N}. This scenario was also supported by the observations taken with the \emph{Insight-Hard X-ray Modulation Telescope} during the initial part of this spin-up episode \citep{2021MNRAS.504.2493L}. On the other hand, low-energy excess could also be explained by scattering in the gas stream around the neutron star and in the stellar wind of the B-type companion star \citep{1996ApJ...463..726S}.

\begin{figure}[hbtp]
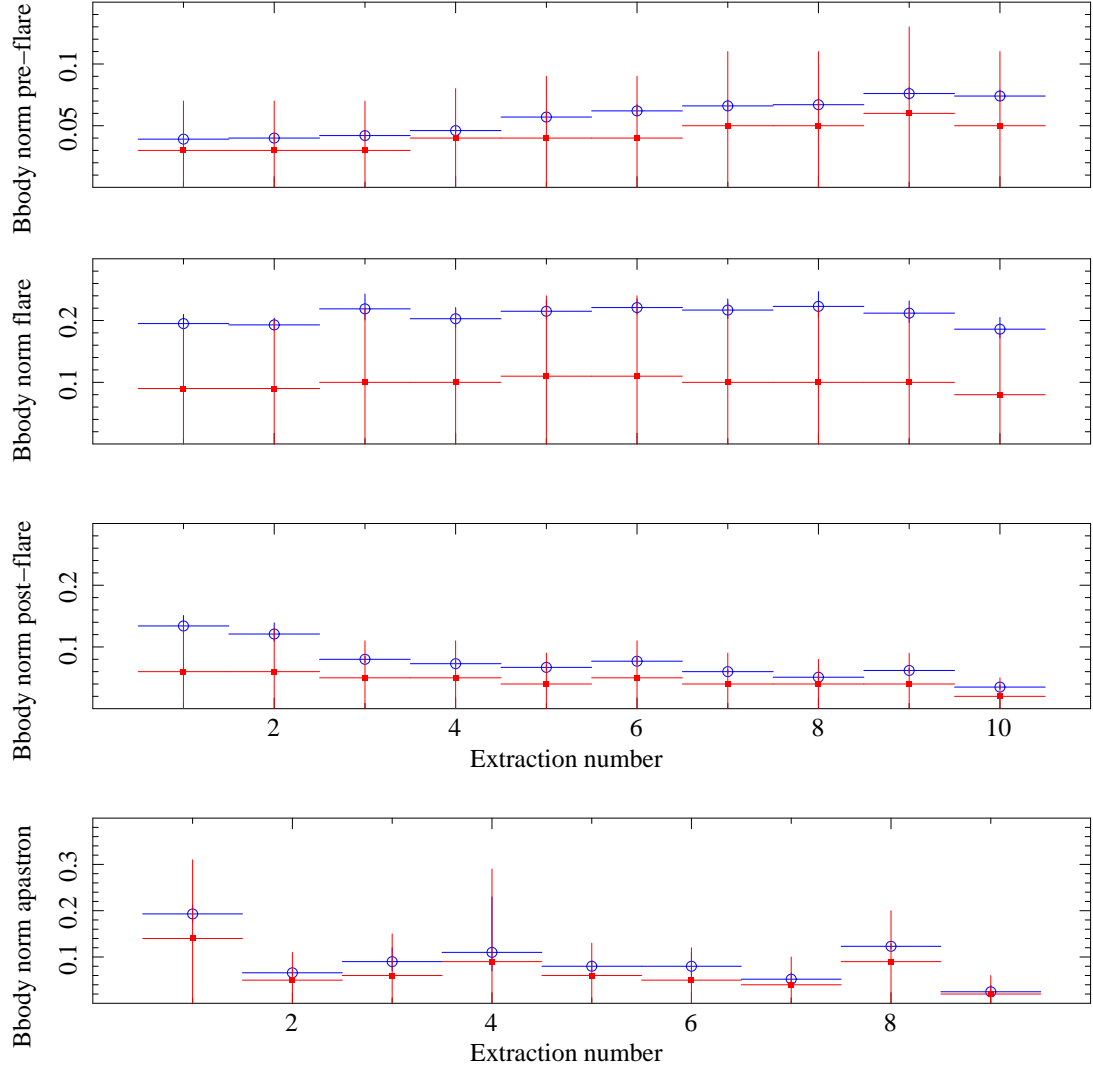

  \centering
  \includegraphics[angle=-90.0,width=\columnwidth]{bbody_norm_inicio.ps} 
  \includegraphics[angle=-90.0,width=\columnwidth]{bbody_norm_fase_intermedia.ps}  
  \includegraphics[angle=-90.0,width=\columnwidth]{bbody_norm_final.ps}
  \includegraphics[angle=-90.0,width=\columnwidth]{bbody_norm_apoastro.ps}
  \caption[]{\label{figure7} %
    Evolution of the \emph{bbody} norm for the pre-periastron flare and apastron outbursts versus the extraction number (\maxi data, model defined by equation~(\ref{ec6})). \emph{Top panel}: pre-flare. \emph{Second panel}: flare. \emph{Third panel}: post-flare. \emph{Bottom panel}: apastron outbursts. Filled red squares: bbody norm calculated by using $L_{39}/D^{2}_{10}$. Open blue circles: \emph{bbody} norm fit value.
  }
\end{figure}

From Figure~\ref{figure7} to Figure~\ref{figure17} the plots show the evolution of the best-fit model parameters during the pre-periastron and apastron passages. In Figures~\ref{figure8}-\ref{figure17} filled black squares represent the pre-flare parameter values, filled red triangles represent the flare parameter values and open blue circles represent the post-flare parameter values.

Taking the uncertainties into account, the results for the \emph{bbody} normalisation (Figure~\ref{figure7}) obtained with \textsc{Xspec} (open blue circles) are consistent with the values obtained using $L_{39}/D^{2}_{10}$ (filled red squares).

The radius of the emission zone (Figure~\ref{figure8}) is of the order of 1 km, which is consistent with a hot spot on the NS surface and with the results of \xmm data. The temperature of the blackbody ($kT$ in keV, Figure~\ref{figure9}) has a constant value between (3.0--3.5) keV in the pre-flare spectra but it has a little enhanced in the sixth extraction which is reflected in a drop of the emission radius (Figure~\ref{figure8}). In the flare spectra, it is in the (5.5--6.5) keV range indicating an increase in the accretion rate which is reflected in a higher X-ray luminosity compared to pre- and post-flare. In the post-flare it is approximately constant, (3.8--4.1) keV, between the third and the tenth extraction, but there is a high increase in the first and second extractions with a value $\approx$ 6.0 keV which is reflected in a clear decrease of the emission zone between $\approx$ (0.2--0.3) km.

\begin{figure}[hbtp]
  \centering
  \includegraphics[angle=-90.0,width=\columnwidth]{radius.ps}
  \includegraphics[angle=-90.0,width=\columnwidth]{radius_apoastro.ps}
  \caption[]{\label{figure8} %
    Evolution of the radius of the emission zone for the pre-periastron flare and apastron outbursts versus the extraction number (\maxi data, model defined by equation~(\ref{ec6})). Filled black squares: pre-flare. Filled red triangles: flare. Open blue circles: post-flare. Filled dark grey stars: apastron outbursts.
  }
\end{figure}

\begin{figure}[hbtp]
  \centering
  \includegraphics[angle=-90.0,width=\columnwidth]{temperatura.ps}
  \includegraphics[angle=-90.0,width=\columnwidth]{temperatura_apoastro.ps}
  \caption[]{\label{figure9} %
    Evolution of the bbody temperature for the pre-periastron flare and apastron outbursts versus the extraction number (\maxi data, model defined by equation~(\ref{ec6})). Filled black squares: pre-flare. Filled red triangles: flare. Open blue circles: post-flare. Filled dark grey stars: apastron outbursts.
  }
\end{figure}

In the flare the unabsorbed flux (Figure~\ref{figure10}) has no significant evolution in the fitting energy range (2.0--20.0) keV, showing that the accretion rate is quite stable. The unabsorbed fluxes show opposite trends between the pre-flare (the flux increases) and the post-flare (the flux decreases) showing that the accretion rate has different behaviour along the pre-periastron passage. A possible explanation for this would be that the absorbing matter is located in the line of sight between the observer and the neutron star in the post-flare ($N_H \gtrsim 2 \times 10^{23}$ cm$^{-2}$) and away from the line of sight in the pre-flare ($N_H < 2 \times 10^{23}$ cm$^{-2}$). Thus, it is confirmed that the distribution of circumstellar matter around the compact object is rather inhomogeneous during the pre-periastron passage \citep{1996ApJ...463..726S,2014MNRAS.441.2539I}.

\begin{figure}[hbtp]
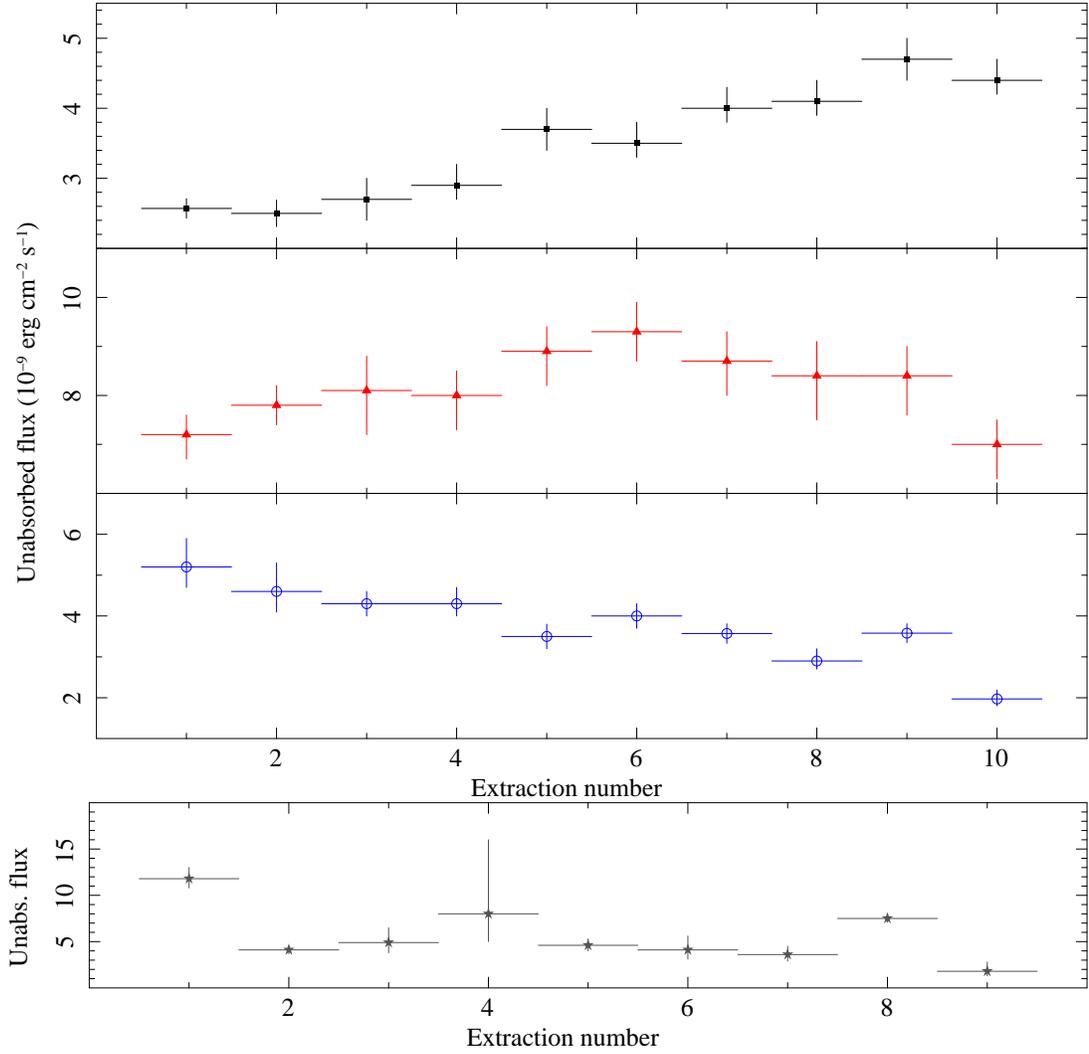

  \centering
  \includegraphics[angle=-90.0,width=\columnwidth]{flujo_no_absorbido.ps}  
  \includegraphics[angle=-90.0,width=\columnwidth]{flujo_no_absorbido_apoastro.ps}
  \caption[]{\label{figure10} %
    Evolution of the unabsorbed flux for the pre-periastron flare and apastron outbursts (in units of 10$^{-9}$ erg cm$^{-2}$ s$^{-1}$) versus the extraction number (\maxi data, model defined by equation~(\ref{ec6})). Filled black squares: pre-flare. Filled red triangles: flare. Open blue circles: post-flare. Filled dark grey stars: apastron outbursts.
  }
\end{figure}

The X-ray luminosity (Figure~\ref{figure11}) is consistent with a constant value ($L_\mathrm{X} \approx 1.3\times10^{37}$ \ergs) in the flare spectra (which agrees with the observations ID 0555200301 and ID 0555200401 from \xmm) and is greater than in the pre-flare ($L_\mathrm{X} = [4-7]\times10^{36}$ \ergs) and in the post-flare ($L_\mathrm{X} = [3-8]\times10^{36}$ \ergs).
This scenario, in which the wind mass-loss rate is insufficient to power the source entirely, indicates that it is needed an additional mechanism to give the neutron star the fuel it needs. Wind-fed \hmxbs are powered by accretion of the radiatively driven wind of the luminous component on the compact object with typical X-ray luminosities $\approx 10^{36}$ \ergs \citep[see][for instance]{2017SSRv..212...59M,2019NewAR..8601546K}, i.e. one order of magnitude smaller than observed in the flare event. A possible mechanism for enhancing X-ray emission is thought to be the presence of a gas stream trailing the neutron star \citep{1996ApJ...463..726S,2014MNRAS.441.2539I}. Another cause could be that an accretion disc has formed around the neutron star \citep{2019A&A...629A.101N,2021MNRAS.504.2493L} or a combination of both possibilities \citep{1996ApJ...463..726S}. Spin-up episodes are usually characterised by an increase in X-ray luminosity, associated with an enhancement of the accreted matter as a consequence of the formation of a temporary accretion disc around the neutron star \citep{2019A&A...629A.101N,2023MNRAS.520.1411M}.

\begin{figure}[hbtp]
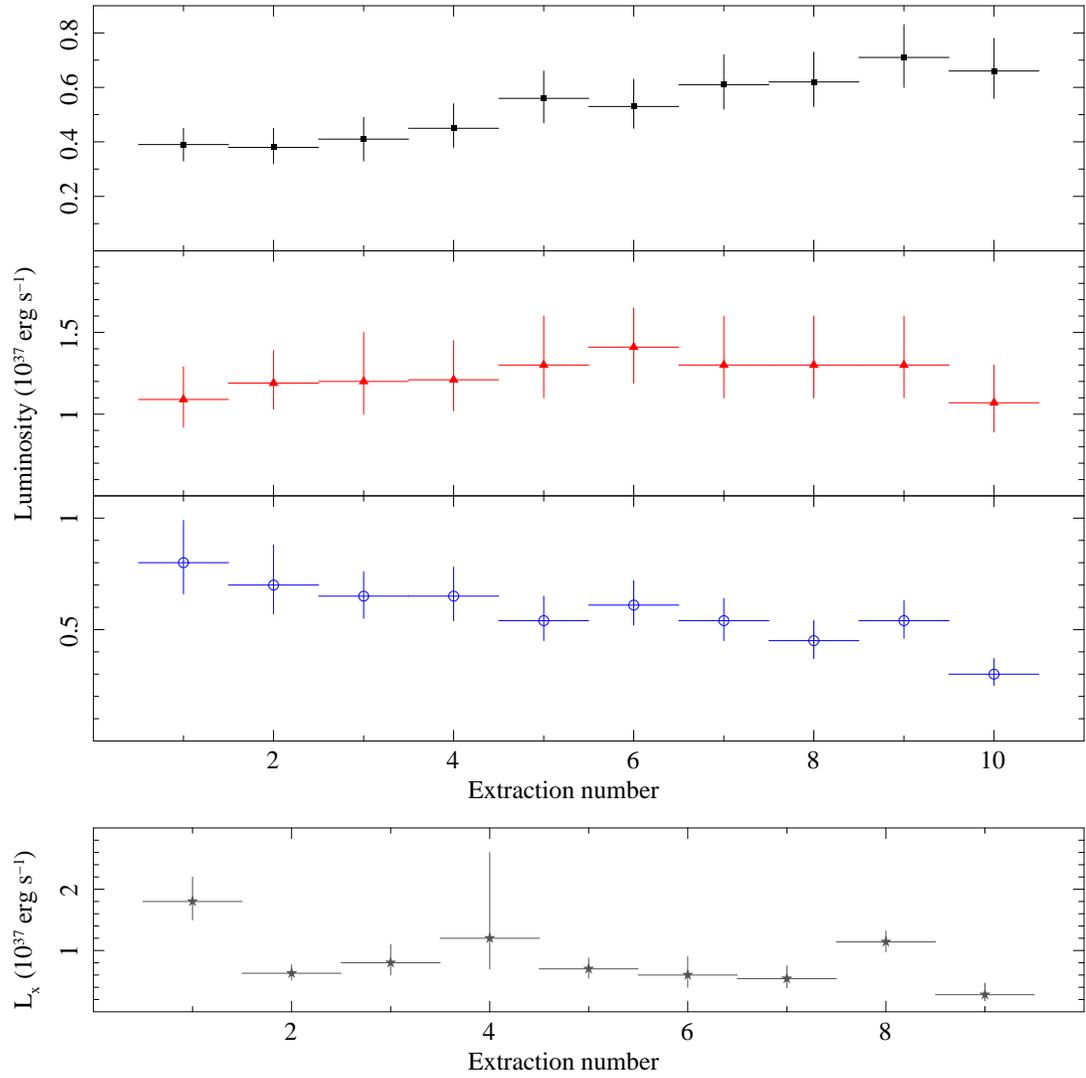

  \centering
  \includegraphics[angle=-90.0,width=\columnwidth]{luminosidad.ps} 
  \includegraphics[angle=-90.0,width=\columnwidth]{luminosidad_apoastro.ps}
  \caption[]{\label{figure11} %
    Evolution of the luminosity for the pre-periastron flare and apastron outbursts versus the extraction number (\maxi data, model defined by equation~(\ref{ec6})). Filled black squares: pre-flare. Filled red triangles: flare. Open blue circles: post-flare. Filled dark grey stars: apastron outbursts.
  }
\end{figure}

This source is seen through a column density in the range $N_H = [10 - 40]\times 10^{22}$ cm$^{-2}$, and all X-rays below $[2-3]$ keV are absorbed. At these column densities a feature due to K edge of Fe at 7.1 keV should be present. Figure~\ref{figure12} shows the optical depth of the iron edge $\left(\tau_\mathrm{Kedge}\right)$ best-fit parameter that lies in the range $[0.3-0.8]$. This represents an ionisation state of nearly neutral iron Fe \textsc{i-v} \citep[if values reported by][are taken into account]{1996ApJ...463..726S,2002ApJ...574..879E,2021MNRAS.501.2522J}.

\begin{figure}[hbtp]
  \centering
  \includegraphics[angle=-90.0,width=\columnwidth]{profundidad_optica.ps}  
  \includegraphics[angle=-90.0,width=\columnwidth]{profundidad_optica_apoastro.ps}
  \caption[]{\label{figure12} %
    Evolution of the optical depth of the edge for the pre-periastron flare and apastron outbursts versus the extraction number (\maxi data, model defined by equation~(\ref{ec6})). Filled black squares: pre-flare. Filled red triangles: flare. Open blue circles: post-flare. Filled dark grey stars: apastron outbursts.
  }
\end{figure}

The long-term averaged spectra in the flare section present a fluorescent iron emission line energy consistent with the Fe K$\alpha$ line and a constant value of $\sim (6.35\pm 0.08)$ keV, as can be seen in Figure~\ref{figure13}, top panel, which implies an ionisation state of iron up to Fe \textsc{xvii}. It was only detected in the first two spectra in the post-flare section and it was totally absent in the pre-flare section. However, \citet{2014MNRAS.441.2539I} performed an orbital phase-resolved spectral analysis of \gx and they reported that the iron emission line was detected in all the orbital phases. The fitted values of the line width, the equivalent width  and the intensity of the Fe K$\alpha$ emission line are plotted in Figure~\ref{figure13}: second, third and bottom panels, respectively. 

\begin{figure}[hbtp]
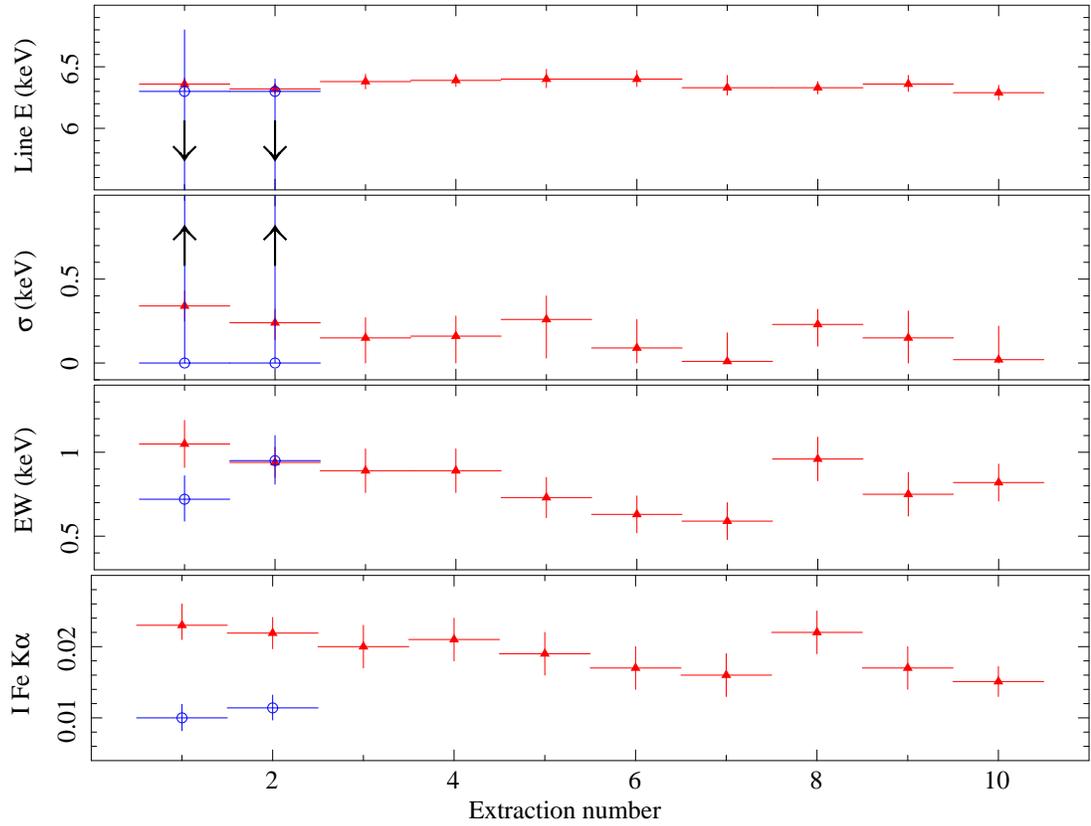

  \centering
  \includegraphics[angle=-90.0,width=\columnwidth]{line_energy.ps} 
  \includegraphics[angle=-90.0,width=\columnwidth]{sigma.ps}
  \includegraphics[angle=-90.0,width=\columnwidth]{anchura_equivalente.ps}
  \includegraphics[angle=-90.0,width=\columnwidth]{intensidad_hierro.ps}
  \caption[]{\label{figure13} %
    The parameters of the Fe K$\alpha$ emission line versus the extraction number (\maxi data, model defined by equation~(\ref{ec6})). \emph{Top panel}: line energy. \emph{Second panel}: line width. \emph{Third panel}: equivalent width. \emph{Bottom panel}: intensity. The unit of the line flux $I$ is photons s$^{-1}$ cm$^{-2}$. Filled red triangles: flare. Open blue circles: post-flare.
  }
\end{figure}

\begin{figure}[hbtp]
  \centering
  \includegraphics[angle=-90.0,width=\columnwidth]{columnas_hidrogeno.ps} 
  \includegraphics[angle=-90.0,width=\columnwidth]{columnas_hidrogeno_apoastro.ps}
  \caption[]{\label{figure17} %
    Evolution of the column density for the pre-periastron flare and apastron outbursts versus the extraction number (\maxi data, model defined by equation~(\ref{ec6})). Filled black squares: pre-flare. Filled red triangles: flare. Open blue circles: post-flare. Filled dark grey stars: apastron outbursts.
  }
\end{figure}

If the emission line luminosity is produced by a thin spherical shell of matter, i.e. a shell of gas surrounding a point source of continuum radiation, with uniform ionisation, composition and density, the line equivalent width should be proportional to the column density \citep{2004ApJS..155..675K,2010ApJ...715..947T} $EW_\mathrm{line} \mathrm{(keV)} \simeq 0.3\, N_H \left(10^{24}\right.$ cm$\left.^{-2}\right)$. The absorbing component varies in the range $\sim [10-30] \times 10^{22}$ cm$^{-2}$ (see second panel from the top of Figure~\ref{figure17}), which implies equivalent width of Fe K$\alpha$ in the range $EW_\mathrm{line} \sim [30-90]$ eV. On the other hand, for an isotropic surrounding cosmic fluorescing plasma, \citet{1985SSRv...40..317I} obtained a relationship between the expected equivalent width of the iron K$\alpha$ line and the hydrogen column density as $EW_\mathrm{expected} \mathrm{(eV)} = 100\, N_H \left(10^{23}\right.$ cm$\left.^{-2}\right)$ that is valid for $N_H < 10^{24}$ cm$^{-2}$ and a photon index of the power-law spectrum of 1.1 \citep[see also][for example]{2002ApJ...574..879E}. They also concluded that \gx corresponds to this case and, therefore, $EW_\mathrm{line} \sim [100-300]$ eV. However, \maxi observations gave equivalent width of the iron emission line in the range $EW_\mathrm{line} \sim [600-1100]$ eV, which implies high deviations from the linear correlation. The highest value of the equivalent width was obtained with the lowest column density. The fact that the Fe K$\alpha$ is not a single line but a superposition of different K$\alpha$ lines of differently strongly ionised iron together with the Compton shoulder at $\sim$6.24 keV and the wind speed could explain these high equivalent widths found in this study \citep{2003ApJ...597L..37W,2011A&A...535A...9F,2021MNRAS.501.2522J}. Taking the uncertainties into account, the best-fit parameters in the pre-periastron orbital phase are consistent with previous studies \citep{2014MNRAS.441.2539I,2023MNRAS.520.1411M}. As a consequence, during the pre-periastron passage the fluorescent iron emission line is not emitted from a spherically symmetric distribution of matter surrounding the neutron star. Although the study has focused on flare episodes, a very small part of the orbit, these results are consistent with those obtained by \citet{2014MNRAS.441.2539I}.

\begin{figure}[hbtp]
  \centering
  \includegraphics[angle=-90.0,width=\columnwidth]{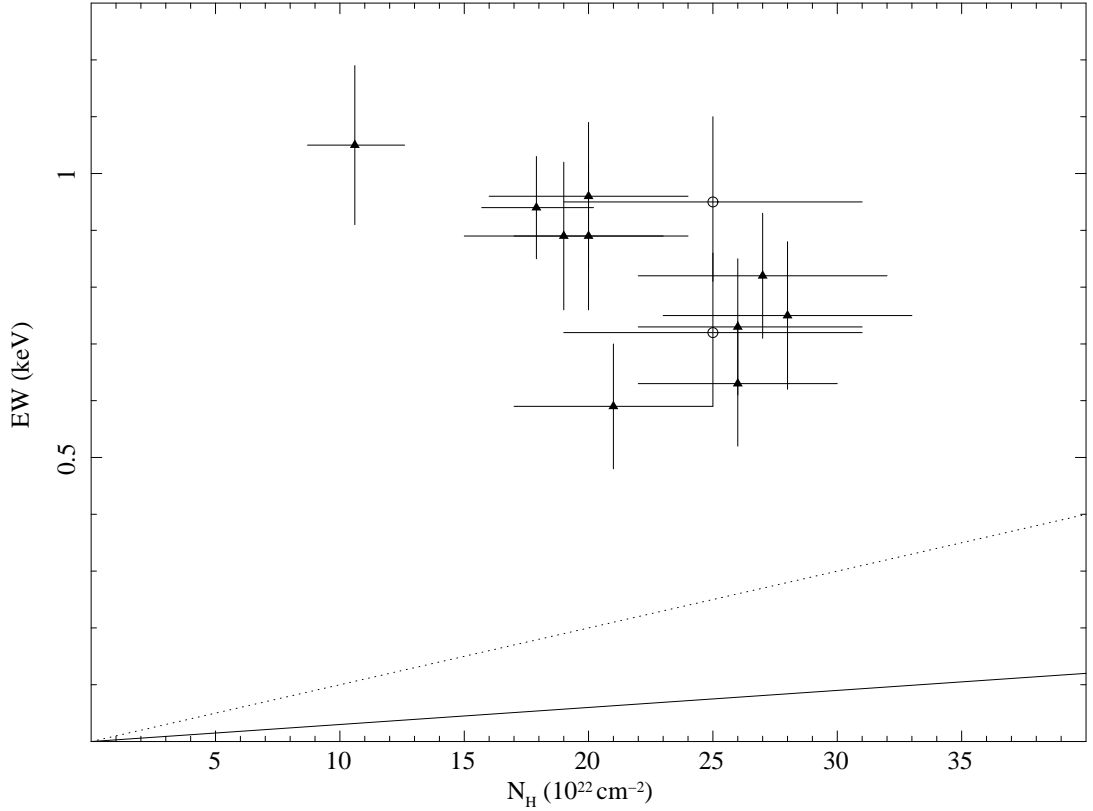} 
  \caption[]{\label{ew_columna_absorcion} %
    Variability of the equivalent width as a function of the column density (also known as the curve of growth). Filled black triangles: flare. Open black circles: post-flare. Dot line represents the relation for an isotropic surrounding cosmic fluorescing plasma, $EW_\mathrm{expected} \mathrm{(eV)} = 100\, N_H \left(10^{23}\right.$ cm$\left.^{-2}\right)$ \citep{1985SSRv...40..317I}. Solid line is the relation for  luminosity which is produced by a thin spherical shell of matter $EW_\mathrm{line} \mathrm{(keV)} \simeq 0.3\, N_H \left(10^{24}\right.$ cm$\left.^{-2}\right)$ \citep{2004ApJS..155..675K,2010ApJ...715..947T}.  
  }
\end{figure}

Figure~\ref{ew_columna_absorcion} shows the equivalent width of the Fe K$\alpha$ against the absorption column density during the pre-periastron passage compared with the theoretical predictions for an isotropically distributed gas and for a spherical shell of gas surrounding the source. In fact, it seems that there is a moderated anti-correlation between them during the pre-periastron passage and it was found a Pearson correlation coefficient  $r = -0.63$. This result is in contrast to other studies such as, \citet{1985SSRv...40..421M}, \citet{2002ApJ...574..879E}, \citet{2011A&A...535A...9F}, \citet{2021MNRAS.501.2522J}, where they found a linear correlation between these two parameters which suggests that the accretion material near the neutron star is spherically distributed. Nevertheless, \citet{2021MNRAS.501.2522J} also found deviations from this linear correlation when the column density was higher than $1.7\times 10^{24}$ cm$^{-2}$. They pointed out that this could be due to the formation of an accretion disc.

\begin{figure}[hbtp]
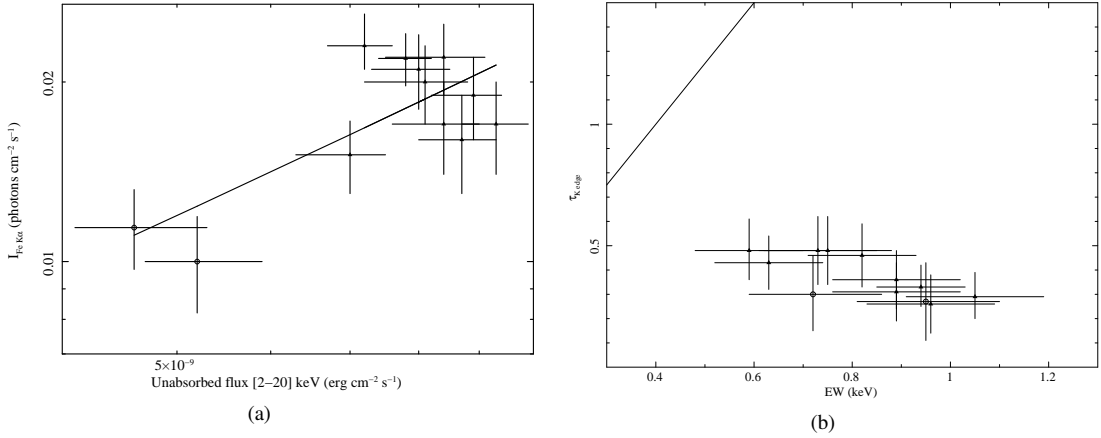

  \centering
  \subfloat[]{%
  \includegraphics[angle=-90.0,width=.49\columnwidth]{unabflux_norm.ps} }
  \subfloat[]{%
  \includegraphics[angle=-90.0,width=.49\columnwidth]{anchura_equivalente_abril_2024.ps} }
  \caption[]{\label{correlated_plots} %
    \emph{Left panel}: Log-log plot of the Fe K$\alpha$ intensity as a function of the unabsorbed flux in the (2-20) keV energy band. The solid line represents a linear fitting. \emph{Right panel}: Plot of the optical depth of the iron K-edge absorption as a function of the equivalent width of the iron K$\alpha$ line (in keV). The solid line is the linear fitting found by \citet{2021MNRAS.501.2522J}. An apparent anti-correlation can be seen during the pre-periastron passage. The larger the EW of the Fe K$\alpha$, the smaller the optical depth of the iron K-edge absorption. Filled black triangles: flare. Open black circles: post-flare.
  }
\end{figure}

In Figure~\ref{correlated_plots} it is shown the intensity of the Fe K$\alpha$ emission line versus the unabsorbed flux of the source in the [2-20] keV energy band (left plot) and the optical depth of the iron K-edge absorption versus the equivalent width of the iron K$\alpha$ line (right plot). It can be seen a moderate linear correlation (Pearson correlation coefficient $r = 0.63$) which should be consistent with the expected line intensities for the hydrogen column densities derived from the model described by equation~\ref{ec6} \citep[see fig.6b in][where the incident spectrum was assumed to be a power-law with the photon index $\Gamma = 0.55$]{1991ApJ...376..245H}. In contrast, from the plot $\tau_\mathrm{edge}$ versus EW, an anticorrelation relationship (Pearson correlation coefficient  $r = -0.75$) seems to be present between these parameters during the pre-periastron passage. The solid line represents the linear fitting found by \citet{2021MNRAS.501.2522J} where they suggested that the reprocessing material reached an optical depth unit for EW $\sim 400$ eV \citep[see also][where they also found a linear correlation]{2010ApJ...715..947T}.

\subsection{Apastron flare spectra}
\label{apastron_flare}

\begin{figure}[htb]
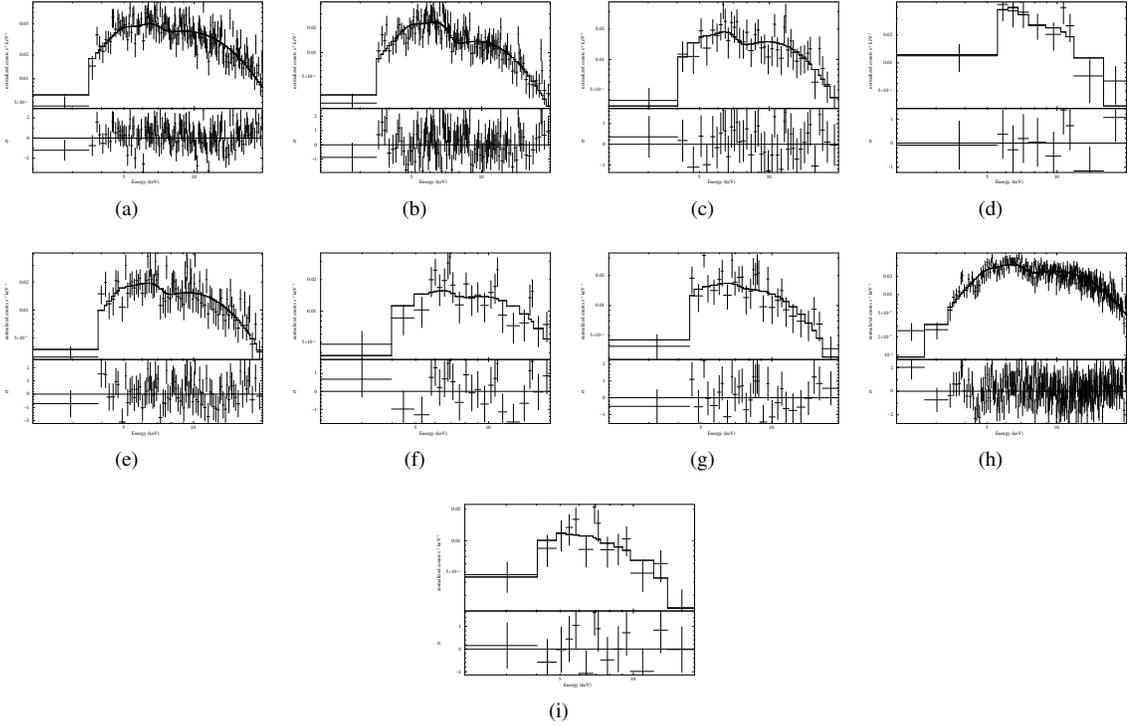

\centering
  \subfloat[]{%
    \includegraphics[angle=-90.0,width=.24\textwidth]{1_apoastro_.ps}}\hfill
  \subfloat[]{%
    \includegraphics[angle=-90.0,width=.24\textwidth]{2_apoastro_.ps}}\hfill
  \subfloat[]{%
    \includegraphics[angle=-90.0,width=.24\textwidth]{3_apoastro_.ps}}\hfill
  \subfloat[]{%
    \includegraphics[angle=-90.0,width=.24\textwidth]{5_apoastro_.ps}}\hfill
  \subfloat[]{%
    \includegraphics[angle=-90.0,width=.24\textwidth]{6_apoastro_.ps}}\hfill
  \subfloat[]{%
    \includegraphics[angle=-90.0,width=.24\textwidth]{7_apoastro_.ps}}\hfill
  \subfloat[]{%
    \includegraphics[angle=-90.0,width=.24\textwidth]{10_apoastro_.ps}}\hfill
  \subfloat[]{%
    \includegraphics[angle=-90.0,width=.24\textwidth]{11_apoastro_.ps}}\hfill
  \subfloat[]{%
    \includegraphics[angle=-90.0,width=.24\textwidth]{12_apoastro_.ps}}\hfill
  \caption{\maxi spectra of \gx in the [2.0--20.0] keV band corresponding to the apastron flare. MJDs from left to right for each spectrum: (a) [55376.5$-$55380.5], (b) [55710.0$-$55713.0], (c) [55870.0.0$-$55873.0], (d) [56163.5$-$56164.5], (e) [57285.0$-$57288.0], (f) [57454.0$-$57457.0], (g) [58029.5$-$58041.5], (h) [58480.5$-$58495.5] and (i) [59119.0$-$59122.0]. Units of the x-axis: Energy (keV). Units of the y-axis: normalized counts s$^{-1}$ keV$^{-1}$.}\label{apastron_spectra}
\end{figure}

Folded light curve of \gx shows two flarelike features at binary orbital phases $\sim 0.26$, i.e. before apastron passage \citep{1991ApJ...376..245H,1996ApJ...463..726S}, and $\sim 0.45$ near-apastron passage \citep[as can be seen in Figure~\ref{figure3}, see also][for instance]{1995ApJ...454..872P,2001ApJ...554..383P}.
In general, periodic near-apastron outbursts show lower intensity than the pre-periastron flare although X-ray emission can sometimes be as low as $\sim 10^{36}$ \ergs, i.e. it cannot be distinguished from the baseline X-ray intensity along the orbit. Thus, the criteria to identify an apastron outburst was that the unabsorbed X-ray flux was greater than $\sim 2\times 10^{-9}$ \ergscm. The orbital phase of the apastron outbursts was also obtained with the parameters reported by \citet{1997ApJ...479..933K}. Then, the Good Time Intervals (GTIs) were derived and the corresponding MJDs are listed in the caption of Figure~\ref{apastron_spectra}. Finally, the spectra were extracted following the process explained in Section~ \ref{pre_periastron_flare}.

In this analysis, all apastron spectra were fitted with the same model as it was used in the pre-periastron flare spectra. The range of the $\chi^{2}_{r}$ values was $[0.7-1.2]$ for all fits. Nevertheless, it should be noted that no Gaussian component has been included here because the fluorescent iron emission line was not detected by \maxi. Figure~\ref{apastron_spectra} shows the data, the absorbed blackbody modified by the iron K-edge absorption best-fit model (top panels), and the residuals for the model (bottom panels). The best-fit model parameters are plotted in Figures~\ref{figure7}-\ref{figure12} and \ref{figure17}, bottom panels.

The unsabsorbed fluxes and the derived radius of the blackbody emission region are shown in Figures~\ref{figure10}) and ~\ref{figure8}, respectively. The results of the radius are consistent with a hot spot on the NS surface \citep{2021MNRAS.501.5892S,2022RMxAA..58..355T} (see Table~\ref{parametros_apoastro}, column 3).

\begin{table}
%\scriptsize
 \caption{Selected fit parameters from apastron flare spectra} 
  \centering
  \begin{tabular}{c c c c c c} \toprule
 Extraction & MJDs & $R_\mathrm{bb}$ & $K_\mathrm{bb}$ & Bbody norm & L$_{X\mathrm{(2.0-20.0\_keV)}}$ \\
 Number &  [Orbital Phase] & (km) & [$L_{39}/D^{2}_{10}$] & fit value & $\left(10^{37} \mathrm{\ergs}\right)$ \\ \midrule
1 &   55376.5$-$55380.5 [0.41--0.51] & $0.79^{+0.20}_{-0.19}$ & $0.14^{+0.17}_{-0.15}$ & $0.193^{+0.019}_{-0.017}$ & $1.8^{+0.4}_{-0.3}$\\ 
2 &  55710.0$-$55713.0 [0.45--0.52] & $0.70^{+0.17}_{-0.16}$ &$0.05^{+0.06}_{-0.05}$ & $0.066^{+0.007}_{-0.006}$ & $0.63^{+0.14}_{-0.11}$ \\
3 & 55870.0$-$55873.0 [0.30--0.37] &  $0.4^{+0.3}_{-0.2}$ &$0.06^{+0.09}_{-0.07}$ & $0.09^{+0.03}_{-0.02}$ & $0.8^{+0.3}_{-0.2}$\\
4 &   56163.5$-$56164.5 [0.38--0.40] & $1.5^{+2.2}_{-1.2}$ &$0.09^{+0.20}_{-0.13}$ & $0.11^{+0.12}_{-0.04}$ & $1.2^{+1.4}_{-0.5}$\\  
5 &  57285.0$-$55288.0 [0.40--0.47] & $0.45^{+0.17}_{-0.14}$ &$0.06^{+0.07}_{-0.06}$ & $0.080^{+0.012}_{-0.010}$ & $0.70^{+0.18}_{-0.15}$\\
6 &  57454.0$-$57457.0 [0.47--0.55] & $0.3^{+0.4}_{-0.3}$ &$0.05^{+0.07}_{-0.06}$ & $0.08^{+0.03}_{-0.02}$ & $0.6^{+0.3}_{-0.2}$\\
7 &  58029.5$-$58041.5 [0.34--0.63] & 0.6$\pm$0.3 &$0.04^{+0.06}_{-0.05}$ & $0.052^{+0.014}_{-0.010}$ & $0.54^{+0.21}_{-0.15}$\\
8 & 58480.5$-$58495.5 [0.21--0.57] & 0.68$\pm$0.11 & $0.09^{+0.11}_{-0.09}$ & $0.123\pm0.006$ & $1.14^{+0.18}_{-0.16}$\\
9 &  59119.0$-$59122.0 [0.60--0.67] & $0.5^{+0.4}_{-0.3}$ &$0.02^{+0.04}_{-0.03}$ & $0.025^{+0.014}_{-0.007}$ & $0.28^{+0.19}_{-0.10}$\\
    \bottomrule    
  \end{tabular}
  \label{parametros_apoastro}
\end{table}

Using the definition of the \emph{bbody normalisation} $K_\mathrm{bb} = L_{39}/D^{2}_{10}$, we have derived its value in the nine spectra (see Table~\ref{parametros_apoastro}, column 4). These values are consistent with those obtained from fitting the model $F(E) = \mathit{tbnew} \times \mathit{bbody} \times \mathit{edge}$, taking the uncertainties into account (see Figure~\ref{figure7}, bottom panel).

The first apastron outburst spectrum observed by \maxi is shown in Figure~\ref{apastron_spectra} (a) whose X-ray luminosity, see Table~\ref{parametros_apoastro}, column 6, $\left(L_{X\mathrm{(a)}} \sim 2\times 10^{37}\right)$ was unusually brighter than pre-periastron flare (see Figure~\ref{figure11} and compare second and bottom panels). This flare was also detected by \fermi and could be associated with a rapid spin-up episode of \gx \citep{2010ATel.2712....1F}. According to the long-term light curve obtained with \fermi, rapid spin-up began on June 23, 2010 (MJD 55\,370.84) and finished on July 22, 2010 (MJD 55\,399.07), as can be seen in Figure~ \ref{spin_frequency_3}. Pulse timing measurements from the interval of June 23.8-July 22.1 showed a spin-up frequency rate of $\dot{\nu}_\mathrm{spin} = \left(3.89\pm 0.08\right)\times 10^{-12}$ Hz/s, giving a spin-up time scale of $\sim 12$ yr. This extracted spectrum covered only 1/7 of the spin-up episode with an spin-up frequency rate of $\dot{\nu}_\mathrm{spin} = \left(6.21\pm 0.21\right)\times 10^{-12}$ Hz/s. The following apastron flare with X-ray luminosity similar to the pre-periastron flare is shown in Figure~\ref{apastron_spectra} (d) and was not associated with any significant changes in the spin period of the neutron star (the spin frequency decreased from $\nu_1 = \left(1.46861\pm 0.00017\right)\times 10^{-3}$ Hz on MJD 55\,611.1160 to $\nu_2 = \left(1.46818\pm 0.00018\right)\times 10^{-3}$ Hz on MJD 55\,616.9968). The last one studied in this work corresponded to the eighth extraction, Figure~\ref{apastron_spectra} (h), where the source presented a new rapid spin-up event \citep{2020ApJ...891...70A,2021MNRAS.504.2493L,2023MNRAS.520.1411M}. In this case, during this observation the spin frequency increased from $\nu_1 = \left(1.467459\pm 0.000011\right)\times 10^{-3}$ Hz on MJD 58\,485.1001 to $\nu_2 = \left(1.472308\pm 0.000007\right)\times 10^{-3}$ Hz on MJD 58\,494.9950 at a rate of $\dot{\nu}_\mathrm{spin} = \left(5.75\pm 0.02\right)\times 10^{-12}$ Hz/s which is near to the spin frequency increased over 18 days (MJD 58\,485.1001--58\,502.9107) $\dot{\nu}_\mathrm{spin} = \left(5.70\pm 0.02\right)\times 10^{-12}$ Hz/s. Then, the spin frequency rose over 24 days (MJD 58\,502.9107--58\,553.1933) at a rate of $\dot{\nu}_\mathrm{spin} = \left(2.37\pm 0.18\right)\times 10^{-12}$ Hz/s.

Another rapid spin-up episode seen with \fermi from MJD 54\,830.9617 to MJD 54\,855.1007 showed a spin-up frequency rate of $\dot{\nu}_\mathrm{spin} = \left(2.97\pm 0.12\right)\times 10^{-12}$ Hz/s \citep{2010ATel.2712....1F}. Moreover, similar spin-up events were detected with the Burst And Transient Source Experiment (BATSE) and reported by \citet{1997ApJ...479..933K,1997ApJS..113..367B}. They found a spin frequency growth over 23 days at a rate of $\dot{\nu}_\mathrm{spin} = 4.5\times 10^{-12}$ Hz/s and over 15 days at a rate of $\dot{\nu}_\mathrm{spin} = 3.0\times 10^{-12}$ Hz/s.

The rest of the apastron flares shown in Figures~\ref{apastron_spectra} (b), (c), (e), (f), (g) and (i) had an X-ray emission slightly lower than a pre-periastron flare which were not associated with any significant changes in the spin period of the neutron star. From MJD 55\,708.9521 to MJD 55\,714.8995, the spin-up rate was $\dot{\nu}_\mathrm{(b)} = \left(1.1\pm 0.6\right)\times 10^{-12}$ Hz/s; from MJD 55\,860 to 55\,875 only two pulse timing measurements were taken, MJD 55\,871.1393 and 55\,873.1066, which gave a spin-down rate of $\dot{\nu}_\mathrm{(c)} = \left(-1.9\pm 1.4\right)\times 10^{-12}$ Hz/s; from MJD 57\,265.2712 to MJD 57\,290.7870, the spin-up rate was $\dot{\nu}_\mathrm{(e)} = \left(9.8\pm 0.8\right)\times 10^{-13}$ Hz/s (although this time interval had a gap without measurements MJD 57\,273--57\,286); from MJD 57\,449.0139 to 57\,458.9447 the spin-down rate was $\dot{\nu}_\mathrm{(f)} = \left(-1.1\pm 0.3\right)\times 10^{-12}$ Hz/s; from MJD 58\,028.9070 to MJD 58\,036.7733, the spin-up rate was $\dot{\nu}_\mathrm{(g)} = \left(1.0\pm 0.7\right)\times 10^{-12}$ Hz/s (but before and after were a gap of 10 days without pulse measurements); and from MJD 59\,117.1258 to MJD 59\,121.0794, the spin-up rate was $\dot{\nu}_\mathrm{(g)} = \left(2.4\pm 1.1\right)\times 10^{-12}$ Hz/s.

As far as it is known from the rapid spin-episodes in \gx, during the apastron passage the source becomes as bright as a pre-periastron flare as a consequence of the formation of a transitory accretion disc. Thus, the mass transfer from the companion star to the neutron star is a rather irregular process during this orbital phase (see X-ray luminosities in Table~\ref{parametros_apoastro}). The strongest events point out that the material is accreted from the stellar wind, possibly from a gas stream \citep{2008MNRAS.384..747L} and probably through a temporary accretion disc \citep{1997ApJ...479..933K,2019A&A...629A.101N,2020ApJ...891...70A,2021MNRAS.504.2493L,2023MNRAS.520.1411M}. However, the fourth apastron flare (see extraction number 4 in Table~\ref{parametros_apoastro} and Figure~\ref{apastron_spectra} (d)) did not show a spin-up of the neutron star and no transitory disc was formed. Therefore, the material should be accreted by the stellar wind and possibly from a gas stream.

It should be noted that the fluorescence iron emission line at $\sim$6.4 keV has not been detected in any spectrum of the apastron flare. Sensitive X-ray observatories such as \asca \citep{2002ApJ...574..879E}, \xmm \citep{2015A&A...576A.108G}, and \chandra \citep{2010ApJ...715..947T} detected and resolved the iron line complex in \gx. In contrast, \maxi does not have enough sensitivity to distinguish between a weak, broad iron K$\alpha$ emission line and a bright X-ray continuum \citep{2015A&A...580A.140R,2022RMxAA..58..355T}. Nevertheless, other long-term orbital phase resolved spectroscopy studies reported the presence of the Fe K$\alpha$ line in all orbital phases \citep{2014MNRAS.441.2539I,2023MNRAS.520.1411M}.

\section{Summary and conclusions}
\label{conclusion}

The main goal of the current study was to determine the long-term variation of \gx in the pre-periastron and apastron flares. The Good Time-Intervals corresponding to these orbital phases were generated using the orbital ephemeris from \citet{1997ApJ...479..933K}.

The main results can be summarise as follows:

-- From the analysis of the \maxi (4.0--10.0) keV light curve, we have estimated the orbital period of the binary system, $P_{\mathrm{orb}} = 41.4 \pm 0.5$ days, being in agreement with the best value derived by
\citet{1997ApJ...479..933K}.

-- Two variations of the model $\mathit{tbnew} \times \mathit{po} + \mathit{tbnew} \times \mathit{po} + \mathit{GL}$  have been applied to find if there was elliptical polarisation due to synchrotron radiation. This model was not able to describe all \maxi data properly.

-- The size of the emitting region on the neutron star surface in the pre-periastron and apastron flares obtained using the \emph{bbody normalisation} (see Figure~\ref{figure8} and Table~\ref{parametros_apoastro}) was compatible with a hot spot. The temperature of the blackbody is in the range [5.1--6.7] keV during the pre-periastron flare and two post-flare but lower than 5.0 keV in the rest of the spectra. It was no clear if there is a connection between the high temperature of the blackbody and the detection of the iron K$\alpha$ line.

-- The X-ray luminosity during the pre-periastron flare was compatible with a constant value ($L_\mathrm{X} \approx 1.3\times10^{37}$ \ergs) indicating an accretion rate quite regular from the stellar wind and a gas stream trailing the neutron star \citep{2008MNRAS.384..747L}. This mass transfer model would also explain the X-ray luminosities obtained in the pre-flare ($L_\mathrm{X} = [4-7]\times 10^{36}$ \ergs), in the post-flare ($L_\mathrm{X} = [3-8]\times 10^{36}$ \ergs) and most of the apastron flares ($L_\mathrm{X} = [3-8]\times 10^{36}$ \ergs). However, two of the strongest apastron flares had an X-ray luminosity comparable to the pre-periastron flare during rare spin-up events. It is believed that a certain amount of angular momentum should be transported through the formation of an accretion disc at this orbital phase \citep{2019A&A...629A.101N,2020ApJ...891...70A,2023MNRAS.520.1411M}. At least one of the largest apastron flares (extraction number 4 in Table~\ref{parametros_apoastro}) was not related to a spin-up episode and therefore it is quite likely that a transient accretion disc did not form. Consequently, it remains unknown why only some apastron flares spin up the X-ray source and on what this depends.

-- The curve of growth showed a moderate anti-correlation between the equivalent width of Fe K$\alpha$ and the column density and clear deviations from spherically distributed absorbing matter \citep{2004ApJS..155..675K,2014MNRAS.441.2539I,2015A&A...576A.108G,2021MNRAS.501.2522J} in the pre-periastron flare. On the other hand, a moderate correlation between the unabsorbed X-ray flux and the intensity of Fe K$\alpha$ was found which could be consistent with the expected values corresponding to the hydrogen column densities in the pre-periastron flare \citep[see Fig. 6b in][]{1991ApJ...376..245H}.

-- The optical depth of the K-edge absorption was moderate anti-correlated to the equivalent width of Fe K$\alpha$ and clearly deviated from the linear correlation reported by \citet{2021MNRAS.501.2522J} \citep[see also the result obtained by][]{2010ApJ...715..947T}.

\begin{acknowledgements}
  
  This research made use of MAXI data provided by RIKEN, JAXA and MAXI team. JJRR acknowledges the support by the Matsumae International Foundation Research Fellowship No14G04, and also thanks the entire MAXI team for the collaboration and hospitality in RIKEN.  We would like to thank particularly T. Mihara and S. Nakahira for invaluable assistance in analysing MAXI data. GSF, JMT \& JJRR acknowledge the financial support from the MCIN with funding from the European Union NextGenerationEU (PRTR-C17.I01) and Generalitat Valenciana Proj. (Athena-XIFU-UA), ref. ASFAE/2022/002. This work has made use of data from the European Space Agency (ESA) mission Gaia (https://www.cosmos.esa.int/gaia), processed by the Gaia Data Processing and Analysis Consortium (DPAC, https://www.cosmos.esa.int/web/gaia/dpac/consortium). Funding for the DPAC has been provided by national institutions, in particular the institutions participating in the Gaia Multilateral Agreement. We thank the anonymous referee for the constructive comments that helped us improve the manuscript.
\end{acknowledgements}

\bibliographystyle{rmaa} % style rmaa.bst
\bibliography{jjrr_ads_v4_} % your references jjrr_ads_v4_

\begin{thebibliography}
\expandafter\ifx\csname natexlab\endcsname\relax\def\natexlab#1{#1}\fi
\expandafter\ifx\csname href\endcsname\relax
  \def\href#1#2{}\fi
\expandafter\ifx\csname urllinklabel\endcsname\relax
  \def\urllinklabel{[LINK]}\fi
\expandafter\ifx\csname adsurllinklabel\endcsname\relax
  \def\adsurllinklabel{[ADS]}\fi

\bibitem[{{Abarr} {et~al.}(2020){Abarr}, {Baring}, {Beheshtipour}, {Beilicke},
  {de Geronimo}, {Dowkontt}, {Errando}, {Guarino}, {Iyer}, {Kislat}, {Kiss},
  {Kitaguchi}, {Krawczynski}, {Lanzi}, {Li}, {Lisalda}, {Okajima}, {Pearce},
  {Press}, {Rauch}, {Stuchlik}, {Takahashi}, {Tang}, {Uchida}, {West}, {Jenke},
  {Krimm}, {Lien}, {Malacaria}, {Miller}, \&
  {Wilson-Hodge}}]{2020ApJ...891...70A}
{Abarr}, Q., {Baring}, M., {Beheshtipour}, B., {Beilicke}, M., {de Geronimo},
  G., {Dowkontt}, P., {Errando}, M., {Guarino}, V., {Iyer}, N., {Kislat}, F.,
  {Kiss}, M., {Kitaguchi}, T., {Krawczynski}, H., {Lanzi}, J., {Li}, S.,
  {Lisalda}, L., {Okajima}, T., {Pearce}, M., {Press}, L., {Rauch}, B.,
  {Stuchlik}, D., {Takahashi}, H., {Tang}, J., {Uchida}, N., {West}, A.,
  {Jenke}, P., {Krimm}, H., {Lien}, A., {Malacaria}, C., {Miller}, J.~M., \&
  {Wilson-Hodge}, C. 2020, \apj, 891, 70


\bibitem[{{Arnaud}(1996)}]{1996ASPC..101...17A}
{Arnaud}, K.~A. in , Astronomical Society of the Pacific Conference Series,
  Vol. 101, Astronomical Data Analysis Software and Systems V, ed. G.~H.
  {Jacoby}J.~{Barnes}, 17


\bibitem[{{Astropy Collaboration} {et~al.}(2022){Astropy Collaboration},
  {Price-Whelan}, {Lim}, {Earl}, {Starkman}, {Bradley}, {Shupe}, {Patil},
  {Corrales}, {Brasseur}, {N{"o}the}, {Donath}, {Tollerud}, {Morris},
  {Ginsburg}, {Vaher}, {Weaver}, {Tocknell}, {Jamieson}, {van Kerkwijk},
  {Robitaille}, {Merry}, {Bachetti}, {G{"u}nther}, {Aldcroft},
  {Alvarado-Montes}, {Archibald}, {B{'o}di}, {Bapat}, {Barentsen}, {Baz{'a}n},
  {Biswas}, {Boquien}, {Burke}, {Cara}, {Cara}, {Conroy}, {Conseil}, {Craig},
  {Cross}, {Cruz}, {D'Eugenio}, {Dencheva}, {Devillepoix}, {Dietrich},
  {Eigenbrot}, {Erben}, {Ferreira}, {Foreman-Mackey}, {Fox}, {Freij}, {Garg},
  {Geda}, {Glattly}, {Gondhalekar}, {Gordon}, {Grant}, {Greenfield}, {Groener},
  {Guest}, {Gurovich}, {Handberg}, {Hart}, {Hatfield-Dodds}, {Homeier},
  {Hosseinzadeh}, {Jenness}, {Jones}, {Joseph}, {Kalmbach}, {Karamehmetoglu},
  {Ka{l}uszy{'n}ski}, {Kelley}, {Kern}, {Kerzendorf}, {Koch}, {Kulumani},
  {Lee}, {Ly}, {Ma}, {MacBride}, {Maljaars}, {Muna}, {Murphy}, {Norman},
  {O'Steen}, {Oman}, {Pacifici}, {Pascual}, {Pascual-Granado}, {Patil},
  {Perren}, {Pickering}, {Rastogi}, {Roulston}, {Ryan}, {Rykoff}, {Sabater},
  {Sakurikar}, {Salgado}, {Sanghi}, {Saunders}, {Savchenko}, {Schwardt},
  {Seifert-Eckert}, {Shih}, {Jain}, {Shukla}, {Sick}, {Simpson},
  {Singanamalla}, {Singer}, {Singhal}, {Sinha}, {Sip{H{o}}cz}, {Spitler},
  {Stansby}, {Streicher}, {{ {S}}umak}, {Swinbank}, {Taranu}, {Tewary},
  {Tremblay}, {Val-Borro}, {Van Kooten}, {Vasovi{'c}}, {Verma}, {de Miranda
  Cardoso}, {Williams}, {Wilson}, {Winkel}, {Wood-Vasey}, {Xue}, {Yoachim},
  {Zhang}, {Zonca}, \& {Astropy Project Contributors}}]{astropy:2022}
{Astropy Collaboration}, {Price-Whelan}, A.~M., {Lim}, P.~L., {Earl}, N.,
  {Starkman}, N., {Bradley}, L., {Shupe}, D.~L., {Patil}, A.~A., {Corrales},
  L., {Brasseur}, C.~E., {N{"o}the}, M., {Donath}, A., {Tollerud}, E.,
  {Morris}, B.~M., {Ginsburg}, A., {Vaher}, E., {Weaver}, B.~A., {Tocknell},
  J., {Jamieson}, W., {van Kerkwijk}, M.~H., {Robitaille}, T.~P., {Merry}, B.,
  {Bachetti}, M., {G{"u}nther}, H.~M., {Aldcroft}, T.~L., {Alvarado-Montes},
  J.~A., {Archibald}, A.~M., {B{'o}di}, A., {Bapat}, S., {Barentsen}, G.,
  {Baz{'a}n}, J., {Biswas}, M., {Boquien}, M., {Burke}, D.~J., {Cara}, D.,
  {Cara}, M., {Conroy}, K.~E., {Conseil}, S., {Craig}, M.~W., {Cross}, R.~M.,
  {Cruz}, K.~L., {D'Eugenio}, F., {Dencheva}, N., {Devillepoix}, H. A.~R.,
  {Dietrich}, J.~P., {Eigenbrot}, A.~D., {Erben}, T., {Ferreira}, L.,
  {Foreman-Mackey}, D., {Fox}, R., {Freij}, N., {Garg}, S., {Geda}, R.,
  {Glattly}, L., {Gondhalekar}, Y., {Gordon}, K.~D., {Grant}, D., {Greenfield},
  P., {Groener}, A.~M., {Guest}, S., {Gurovich}, S., {Handberg}, R., {Hart},
  A., {Hatfield-Dodds}, Z., {Homeier}, D., {Hosseinzadeh}, G., {Jenness}, T.,
  {Jones}, C.~K., {Joseph}, P., {Kalmbach}, J.~B., {Karamehmetoglu}, E.,
  {Ka{l}uszy{'n}ski}, M., {Kelley}, M. S.~P., {Kern}, N., {Kerzendorf}, W.~E.,
  {Koch}, E.~W., {Kulumani}, S., {Lee}, A., {Ly}, C., {Ma}, Z., {MacBride}, C.,
  {Maljaars}, J.~M., {Muna}, D., {Murphy}, N.~A., {Norman}, H., {O'Steen}, R.,
  {Oman}, K.~A., {Pacifici}, C., {Pascual}, S., {Pascual-Granado}, J., {Patil},
  R.~R., {Perren}, G.~I., {Pickering}, T.~E., {Rastogi}, T., {Roulston}, B.~R.,
  {Ryan}, D.~F., {Rykoff}, E.~S., {Sabater}, J., {Sakurikar}, P., {Salgado},
  J., {Sanghi}, A., {Saunders}, N., {Savchenko}, V., {Schwardt}, L.,
  {Seifert-Eckert}, M., {Shih}, A.~Y., {Jain}, A.~S., {Shukla}, G., {Sick}, J.,
  {Simpson}, C., {Singanamalla}, S., {Singer}, L.~P., {Singhal}, J., {Sinha},
  M., {Sip{H{o}}cz}, B.~M., {Spitler}, L.~R., {Stansby}, D., {Streicher}, O.,
  {{ {S}}umak}, J., {Swinbank}, J.~D., {Taranu}, D.~S., {Tewary}, N.,
  {Tremblay}, G.~R., {Val-Borro}, M.~d., {Van Kooten}, S.~J., {Vasovi{'c}}, Z.,
  {Verma}, S., {de Miranda Cardoso}, J.~V., {Williams}, P. K.~G., {Wilson},
  T.~J., {Winkel}, B., {Wood-Vasey}, W.~M., {Xue}, R., {Yoachim}, P., {Zhang},
  C., {Zonca}, A., \& {Astropy Project Contributors}. 2022, \apj, 935, 167


\bibitem[{{Bailer-Jones} {et~al.}(2021){Bailer-Jones}, {Rybizki}, {Fouesneau},
  {Demleitner}, \& {Andrae}}]{2021AJ....161..147B}
{Bailer-Jones}, C.~A.~L., {Rybizki}, J., {Fouesneau}, M., {Demleitner}, M., \&
  {Andrae}, R. 2021, \aj, 161, 147


\bibitem[{{Bildsten} {et~al.}(1997){Bildsten}, {Chakrabarty}, {Chiu}, {Finger},
  {Koh}, {Nelson}, {Prince}, {Rubin}, {Scott}, {Stollberg}, {Vaughan},
  {Wilson}, \& {Wilson}}]{1997ApJS..113..367B}
{Bildsten}, L., {Chakrabarty}, D., {Chiu}, J., {Finger}, M.~H., {Koh}, D.~T.,
  {Nelson}, R.~W., {Prince}, T.~A., {Rubin}, B.~C., {Scott}, D.~M.,
  {Stollberg}, M., {Vaughan}, B.~A., {Wilson}, C.~A., \& {Wilson}, R.~B. 1997,
  \apjs, 113, 367


\bibitem[{{Ebisawa} {et~al.}(1996){Ebisawa}, {Day}, {Kallman}, {Nagase},
  {Kotani}, {Kawashima}, {Kitamoto}, \& {Woo}}]{1996PASJ...48..425E}
{Ebisawa}, K., {Day}, C. S.~R., {Kallman}, T.~R., {Nagase}, F., {Kotani}, T.,
  {Kawashima}, K., {Kitamoto}, S., \& {Woo}, J.~W. 1996, \pasj, 48, 425


\bibitem[{{Endo} {et~al.}(2002){Endo}, {Ishida}, {Masai}, {Kunieda}, {Inoue},
  \& {Nagase}}]{2002ApJ...574..879E}
{Endo}, T., {Ishida}, M., {Masai}, K., {Kunieda}, H., {Inoue}, H., \& {Nagase},
  F. 2002, \apj, 574, 879


\bibitem[{{Finger} {et~al.}(2009){Finger}, {Beklen}, {Narayana Bhat},
  {Paciesas}, {Connaughton}, {Buckley}, {Camero-Arranz}, {Coe}, {Jenke},
  {Kanbach}, {Negueruela}, \& {Wilson-Hodge}}]{2009arXiv0912.3847F}
{Finger}, M.~H., {Beklen}, E., {Narayana Bhat}, P., {Paciesas}, W.~S.,
  {Connaughton}, V., {Buckley}, D.~A.~H., {Camero-Arranz}, A., {Coe}, M.~J.,
  {Jenke}, P., {Kanbach}, G., {Negueruela}, I., \& {Wilson-Hodge}, C.~A. 2009,
  ArXiv e-prints


\bibitem[{{Finger} {et~al.}(2010){Finger}, {Camero-Arranz}, {Wilson-Hodge}, \&
  {Jenke}}]{2010ATel.2712....1F}
{Finger}, M.~H., {Camero-Arranz}, A., {Wilson-Hodge}, C., \& {Jenke}, P. 2010,
  The Astronomer's Telegram, 2712, 1


\bibitem[{{F{\"u}rst} {et~al.}(2011){F{\"u}rst}, {Suchy}, {Kreykenbohm},
  {Barrag{\'a}n}, {Wilms}, {Pottschmidt}, {Caballero}, {Kretschmar},
  {Ferrigno}, \& {Rothschild}}]{2011A&A...535A...9F}
{F{\"u}rst}, F., {Suchy}, S., {Kreykenbohm}, I., {Barrag{\'a}n}, L., {Wilms},
  J., {Pottschmidt}, K., {Caballero}, I., {Kretschmar}, P., {Ferrigno}, C., \&
  {Rothschild}, R.~E. 2011, \aap, 535, A9


\bibitem[{Gim\'enez Ca\~{n}ete \& Castro~Tirado(2005)}]{astronomia_2005}
Gim\'enez Ca\~{n}ete, A. \& Castro~Tirado, A. 2005, Astronom\'ia X (Equipo
  Sirius)


\bibitem[{{Gim{\'e}nez-Garc{\'\i}a} {et~al.}(2015){Gim{\'e}nez-Garc{\'\i}a},
  {Torrej{\'o}n}, {Eikmann}, {Mart{\'\i}nez-N{\'u}{\~n}ez}, {Oskinova},
  {Rodes-Roca}, \& {Bernab{\'e}u}}]{2015A&A...576A.108G}
{Gim{\'e}nez-Garc{\'\i}a}, A., {Torrej{\'o}n}, J.~M., {Eikmann}, W.,
  {Mart{\'\i}nez-N{\'u}{\~n}ez}, S., {Oskinova}, L.~M., {Rodes-Roca}, J.~J., \&
  {Bernab{\'e}u}, G. 2015, \aap, 576, A108


\bibitem[{{Haberl}(1991{\natexlab{a}})}]{1991A&A...252..272H}
{Haberl}, F. 1991{\natexlab{a}}, \aap, 252, 272


\bibitem[{{Haberl}(1991{\natexlab{b}})}]{1991ApJ...376..245H}
---. 1991{\natexlab{b}}, \apj, 376, 245


\bibitem[{{Han} \& {Demir}(2009)}]{2009PhRvA..80e2503H}
{Han}, I. \& {Demir}, L. 2009, \pra, 80, 052503


\bibitem[{{Inoue}(1985)}]{1985SSRv...40..317I}
{Inoue}, H. 1985, \ssr, 40, 317


\bibitem[{{Islam} \& {Paul}(2014)}]{2014MNRAS.441.2539I}
{Islam}, N. \& {Paul}, B. 2014, \mnras, 441, 2539


\bibitem[{{Ji} {et~al.}(2021){Ji}, {Doroshenko}, {Suleimanov}, {Santangelo},
  {Orlandini}, {Liu}, {Ducci}, {Zhang}, {Nabizadeh}, {Gavran}, {Zhang}, {Ge},
  {Li}, {Tao}, {Bu}, {Qu}, {Lu}, {Chen}, {Song}, {Li}, {Xu}, {Cao}, {Chen},
  {Liu}, {Cai}, {Chang}, {Chen}, {Chen}, {Cui}, {Du}, {Gao}, {Gao}, {Gu},
  {Guan}, {Guo}, {Han}, {Huang}, {Huo}, {Jia}, {Jiang}, {Jin}, {Kong}, {Li},
  {Li}, {Li}, {Li}, {Li}, {Li}, {Li}, {Liang}, {Liao}, {Liu}, {Liu}, {Liu},
  {Liu}, {Lu}, {Luo}, {Luo}, {Ma}, {Ma}, {Meng}, {Nang}, {Nie}, {Ou}, {Ren},
  {Sai}, {Song}, {Sun}, {Tan}, {Tuo}, {Wang}, {Wang}, {Wang}, {Wang}, {Wang},
  {Wen}, {Wu}, {Wu}, {Wu}, {Xiao}, {Xiao}, {Xiong}, {Yang}, {Yang}, {Yang},
  {Yang}, {Yi}, {Yin}, {You}, {Zhang}, {Zhang}, {Zhang}, {Zhang}, {Zhang},
  {Zhang}, {Zhang}, {Zhang}, {Zhang}, {Zhao}, {Zhao}, {Zheng}, {Zheng}, \&
  {Zhou}}]{2021MNRAS.501.2522J}
{Ji}, L., {Doroshenko}, V., {Suleimanov}, V., {Santangelo}, A., {Orlandini},
  M., {Liu}, J., {Ducci}, L., {Zhang}, S.~N., {Nabizadeh}, A., {Gavran}, D.,
  {Zhang}, S., {Ge}, M.~Y., {Li}, X.~B., {Tao}, L., {Bu}, Q.~C., {Qu}, J.~L.,
  {Lu}, F.~J., {Chen}, L., {Song}, L.~M., {Li}, T.~P., {Xu}, Y.~P., {Cao},
  X.~L., {Chen}, Y., {Liu}, C.~Z., {Cai}, C., {Chang}, Z., {Chen}, T.~X.,
  {Chen}, Y.~P., {Cui}, W.~W., {Du}, Y.~Y., {Gao}, G.~H., {Gao}, H., {Gu},
  Y.~D., {Guan}, J., {Guo}, C.~C., {Han}, D.~W., {Huang}, Y., {Huo}, J., {Jia},
  S.~M., {Jiang}, W.~C., {Jin}, J., {Kong}, L.~D., {Li}, B., {Li}, C.~K., {Li},
  G., {Li}, W., {Li}, X., {Li}, X.~F., {Li}, Z.~W., {Liang}, X.~H., {Liao},
  J.~Y., {Liu}, B.~S., {Liu}, H.~X., {Liu}, H.~W., {Liu}, X.~J., {Lu}, X.~F.,
  {Luo}, Q., {Luo}, T., {Ma}, R.~C., {Ma}, X., {Meng}, B., {Nang}, Y., {Nie},
  J.~Y., {Ou}, G., {Ren}, X.~Q., {Sai}, N., {Song}, X.~Y., {Sun}, L., {Tan},
  Y., {Tuo}, Y.~L., {Wang}, C., {Wang}, L.~J., {Wang}, P.~J., {Wang}, W.~S.,
  {Wang}, Y.~S., {Wen}, X.~Y., {Wu}, B.~Y., {Wu}, B.~B., {Wu}, M., {Xiao},
  G.~C., {Xiao}, S., {Xiong}, S.~L., {Yang}, R.~J., {Yang}, S., {Yang}, Y.-J.,
  {Yang}, Y.-J., {Yi}, Q.~B., {Yin}, Q.~Q., {You}, Y., {Zhang}, F., {Zhang},
  H.~M., {Zhang}, J., {Zhang}, P., {Zhang}, W., {Zhang}, W.~C., {Zhang}, Y.,
  {Zhang}, Y.~F., {Zhang}, Y.~H., {Zhao}, H.~S., {Zhao}, X.~F., {Zheng}, S.~J.,
  {Zheng}, Y.~G., \& {Zhou}, D.~K. 2021, \mnras, 501, 2522


\bibitem[{{Kallman} {et~al.}(2004){Kallman}, {Palmeri}, {Bautista}, {Mendoza},
  \& {Krolik}}]{2004ApJS..155..675K}
{Kallman}, T.~R., {Palmeri}, P., {Bautista}, M.~A., {Mendoza}, C., \& {Krolik},
  J.~H. 2004, \apjs, 155, 675


\bibitem[{{Kaper} {et~al.}(1995){Kaper}, {Lamers}, {Ruymaekers}, {van den
  Heuvel}, \& {Zuiderwijk}}]{1995A&A...300..446K}
{Kaper}, L., {Lamers}, H.~J.~G.~L.~M., {Ruymaekers}, E., {van den Heuvel},
  E.~P.~J., \& {Zuiderwijk}, E.~J. 1995, \aap, 300, 446


\bibitem[{{Kaper} {et~al.}(2006){Kaper}, {van der Meer}, \&
  {Najarro}}]{2006A&A...457..595K}
{Kaper}, L., {van der Meer}, A., \& {Najarro}, F. 2006, \aap, 457, 595


\bibitem[{{Koh} {et~al.}(1997){Koh}, {Bildsten}, {Chakrabarty}, {Nelson},
  {Prince}, {Vaughan}, {Finger}, {Wilson}, \& {Rubin}}]{1997ApJ...479..933K}
{Koh}, D.~T., {Bildsten}, L., {Chakrabarty}, D., {Nelson}, R.~W., {Prince},
  T.~A., {Vaughan}, B.~A., {Finger}, M.~H., {Wilson}, R.~B., \& {Rubin}, B.~C.
  1997, \apj, 479, 933


\bibitem[{{Kretschmar} {et~al.}(2019){Kretschmar}, {F{\"u}rst}, {Sidoli},
  {Bozzo}, {Alfonso-Garz{\'o}n}, {Bodaghee}, {Chaty}, {Chernyakova},
  {Ferrigno}, {Manousakis}, {Negueruela}, {Postnov}, {Paizis}, {Reig},
  {Rodes-Roca}, {Tsygankov}, {Bird}, {Bissinger n{\'e} K{\"u}hnel}, {Blay},
  {Caballero}, {Coe}, {Domingo}, {Doroshenko}, {Ducci}, {Falanga}, {Grebenev},
  {Grinberg}, {Hemphill}, {Kreykenbohm}, {Kreykenbohm n{\'e} Fritz}, {Li},
  {Lutovinov}, {Mart{\'\i}nez-N{\'u}{\~n}ez}, {Mas-Hesse}, {Masetti},
  {McBride}, {Neronov}, {Pottschmidt}, {Rodriguez}, {Romano}, {Rothschild},
  {Santangelo}, {Sguera}, {Staubert}, {Tomsick}, {Torrej{\'o}n}, {Torres},
  {Walter}, {Wilms}, {Wilson-Hodge}, \& {Zhang}}]{2019NewAR..8601546K}
{Kretschmar}, P., {F{\"u}rst}, F., {Sidoli}, L., {Bozzo}, E.,
  {Alfonso-Garz{\'o}n}, J., {Bodaghee}, A., {Chaty}, S., {Chernyakova}, M.,
  {Ferrigno}, C., {Manousakis}, A., {Negueruela}, I., {Postnov}, K., {Paizis},
  A., {Reig}, P., {Rodes-Roca}, J.~J., {Tsygankov}, S., {Bird}, A.~J.,
  {Bissinger n{\'e} K{\"u}hnel}, M., {Blay}, P., {Caballero}, I., {Coe}, M.~J.,
  {Domingo}, A., {Doroshenko}, V., {Ducci}, L., {Falanga}, M., {Grebenev},
  S.~A., {Grinberg}, V., {Hemphill}, P., {Kreykenbohm}, I., {Kreykenbohm n{\'e}
  Fritz}, S., {Li}, J., {Lutovinov}, A.~A., {Mart{\'\i}nez-N{\'u}{\~n}ez}, S.,
  {Mas-Hesse}, J.~M., {Masetti}, N., {McBride}, V.~A., {Neronov}, A.,
  {Pottschmidt}, K., {Rodriguez}, J., {Romano}, P., {Rothschild}, R.~E.,
  {Santangelo}, A., {Sguera}, V., {Staubert}, R., {Tomsick}, J.~A.,
  {Torrej{\'o}n}, J.~M., {Torres}, D.~F., {Walter}, R., {Wilms}, J.,
  {Wilson-Hodge}, C.~A., \& {Zhang}, S. 2019, \nar, 86, 101546


\bibitem[{{La Barbera} {et~al.}(2005){La Barbera}, {Segreto}, {Santangelo},
  {Kreykenbohm}, \& {Orlandini}}]{2005A&A...438..617L}
{La Barbera}, A., {Segreto}, A., {Santangelo}, A., {Kreykenbohm}, I., \&
  {Orlandini}, M. 2005, \aap, 438, 617


\bibitem[{{Leahy}(1991)}]{1991MNRAS.250..310L}
{Leahy}, D.~A. 1991, \mnras, 250, 310


\bibitem[{{Leahy} \& {Kostka}(2008)}]{2008MNRAS.384..747L}
{Leahy}, D.~A. \& {Kostka}, M. 2008, \mnras, 384, 747


\bibitem[{{Leahy} {et~al.}(1989){Leahy}, {Matsuoka}, {Kawai}, \&
  {Makino}}]{1989MNRAS.236..603L}
{Leahy}, D.~A., {Matsuoka}, M., {Kawai}, N., \& {Makino}, F. 1989, \mnras, 236,
  603


\bibitem[{{Leahy} {et~al.}(1988){Leahy}, {Nakajo}, {Matsuoka}, {Kawai},
  {Koyama}, \& {Makino}}]{1988PASJ...40..197L}
{Leahy}, D.~A., {Nakajo}, M., {Matsuoka}, M., {Kawai}, N., {Koyama}, K., \&
  {Makino}, F. 1988, \pasj, 40, 197


\bibitem[{{Lewin} {et~al.}(1971){Lewin}, {McClintock}, {Ryckman}, \&
  {Smith}}]{1971ApJ...166L..69L}
{Lewin}, W.~H.~G., {McClintock}, J.~E., {Ryckman}, S.~G., \& {Smith}, W.~B.
  1971, \apjl, 166, L69


\bibitem[{{Liu} {et~al.}(2021){Liu}, {Ji}, {Jenke}, {Doroshenko}, {Liao}, {Li},
  {Zhang}, {Orlandini}, {Ge}, {Zhang}, \& {Santangelo}}]{2021MNRAS.504.2493L}
{Liu}, J., {Ji}, L., {Jenke}, P.~A., {Doroshenko}, V., {Liao}, Z., {Li}, X.,
  {Zhang}, S., {Orlandini}, M., {Ge}, M., {Zhang}, S., \& {Santangelo}, A.
  2021, \mnras, 504, 2493


\bibitem[{{Makino} {et~al.}(1985){Makino}, {Leahy}, \&
  {Kawai}}]{1985SSRv...40..421M}
{Makino}, F., {Leahy}, D.~A., \& {Kawai}, N. 1985, \ssr, 40, 421


\bibitem[{{Malacaria} {et~al.}(2020){Malacaria}, {Jenke}, {Roberts},
  {Wilson-Hodge}, {Cleveland}, {Mailyan}, \& {GBM Accreting Pulsars Program
  Team}}]{2020ApJ...896...90M}
{Malacaria}, C., {Jenke}, P., {Roberts}, O.~J., {Wilson-Hodge}, C.~A.,
  {Cleveland}, W.~H., {Mailyan}, B., \& {GBM Accreting Pulsars Program Team}.
  2020, \apj, 896, 90


\bibitem[{{Manikantan} {et~al.}(2023){Manikantan}, {Paul}, {Roy}, \&
  {Rana}}]{2023MNRAS.520.1411M}
{Manikantan}, H., {Paul}, B., {Roy}, K., \& {Rana}, V. 2023, \mnras, 520, 1411


\bibitem[{{Mart{\'\i}nez-N{\'u}{\~n}ez}
  {et~al.}(2017){Mart{\'\i}nez-N{\'u}{\~n}ez}, {Kretschmar}, {Bozzo},
  {Oskinova}, {Puls}, {Sidoli}, {Sundqvist}, {Blay}, {Falanga}, {F{\"u}rst},
  {G{\'\i}menez-Garc{\'\i}a}, {Kreykenbohm}, {K{\"u}hnel}, {Sander},
  {Torrej{\'o}n}, \& {Wilms}}]{2017SSRv..212...59M}
{Mart{\'\i}nez-N{\'u}{\~n}ez}, S., {Kretschmar}, P., {Bozzo}, E., {Oskinova},
  L.~M., {Puls}, J., {Sidoli}, L., {Sundqvist}, J.~O., {Blay}, P., {Falanga},
  M., {F{\"u}rst}, F., {G{\'\i}menez-Garc{\'\i}a}, A., {Kreykenbohm}, I.,
  {K{\"u}hnel}, M., {Sander}, A., {Torrej{\'o}n}, J.~M., \& {Wilms}, J. 2017,
  \ssr, 212, 59


\bibitem[{{McClintock} {et~al.}(1971){McClintock}, {Ricker}, \&
  {Lewin}}]{1971ApJ...166L..73M}
{McClintock}, J.~E., {Ricker}, G.~R., \& {Lewin}, W.~H.~G. 1971, \apjl, 166,
  L73


\bibitem[{{Mukherjee} \& {Paul}(2004)}]{2004A&A...427..567M}
{Mukherjee}, U. \& {Paul}, B. 2004, \aap, 427, 567


\bibitem[{{Nabizadeh} {et~al.}(2019){Nabizadeh}, {M{\"o}nkk{\"o}nen},
  {Tsygankov}, {Doroshenko}, {Molkov}, \& {Poutanen}}]{2019A&A...629A.101N}
{Nabizadeh}, A., {M{\"o}nkk{\"o}nen}, J., {Tsygankov}, S.~S., {Doroshenko}, V.,
  {Molkov}, S.~V., \& {Poutanen}, J. 2019, \aap, 629, A101


\bibitem[{{Pravdo} {et~al.}(1995){Pravdo}, {Day}, {Angelini}, {Harmon},
  {Yoshida}, \& {Saraswat}}]{1995ApJ...454..872P}
{Pravdo}, S.~H., {Day}, C.~S.~R., {Angelini}, L., {Harmon}, B.~A., {Yoshida},
  A., \& {Saraswat}, P. 1995, \apj, 454, 872


\bibitem[{{Pravdo} \& {Ghosh}(2001)}]{2001ApJ...554..383P}
{Pravdo}, S.~H. \& {Ghosh}, P. 2001, \apj, 554, 383


\bibitem[{{Press} \& {Rybicki}(1989)}]{1989ApJ...338..277P}
{Press}, W.~H. \& {Rybicki}, G.~B. 1989, \apj, 338, 277


\bibitem[{{Rodes-Roca} {et~al.}(2015){Rodes-Roca}, {Mihara}, {Nakahira},
  {Torrej{\'o}n}, {Gim{\'e}nez-Garc{\'\i}a}, \&
  {Bernab{\'e}u}}]{2015A&A...580A.140R}
{Rodes-Roca}, J.~J., {Mihara}, T., {Nakahira}, S., {Torrej{\'o}n}, J.~M.,
  {Gim{\'e}nez-Garc{\'\i}a}, {\'A}., \& {Bernab{\'e}u}, G. 2015, \aap, 580,
  A140


\bibitem[{{Roy} {et~al.}(2024){Roy}, {Manikantan}, \&
  {Paul}}]{2024MNRAS.527.2652R}
{Roy}, K., {Manikantan}, H., \& {Paul}, B. 2024, \mnras, 527, 2652


\bibitem[{{Sanjurjo-Ferr{\'\i}n} {et~al.}(2021){Sanjurjo-Ferr{\'\i}n},
  {Torrej{\'o}n}, {Postnov}, {Oskinova}, {Rodes-Roca}, \&
  {Bernabeu}}]{2021MNRAS.501.5892S}
{Sanjurjo-Ferr{\'\i}n}, G., {Torrej{\'o}n}, J.~M., {Postnov}, K., {Oskinova},
  L., {Rodes-Roca}, J.~J., \& {Bernabeu}, G. 2021, \mnras, 501, 5892


\bibitem[{{Saraswat} {et~al.}(1996){Saraswat}, {Yoshida}, {Mihara}, {Kawai},
  {Takeshima}, {Nagase}, {Makishima}, {Tashiro}, {Leahy}, {Pravdo}, {Day}, \&
  {Angelini}}]{1996ApJ...463..726S}
{Saraswat}, P., {Yoshida}, A., {Mihara}, T., {Kawai}, N., {Takeshima}, T.,
  {Nagase}, F., {Makishima}, K., {Tashiro}, M., {Leahy}, D.~A., {Pravdo}, S.,
  {Day}, C.~S.~R., \& {Angelini}, L. 1996, \apj, 463, 726


\bibitem[{{Sato} {et~al.}(1986){Sato}, {Nagase}, {Kawai}, {Kelley},
  {Rappaport}, \& {White}}]{1986ApJ...304..241S}
{Sato}, N., {Nagase}, F., {Kawai}, N., {Kelley}, R.~L., {Rappaport}, S., \&
  {White}, N.~E. 1986, \apj, 304, 241


\bibitem[{{Stevens}(1988)}]{1988MNRAS.232..199S}
{Stevens}, I.~R. 1988, \mnras, 232, 199


\bibitem[{{Suchy} {et~al.}(2012){Suchy}, {F{\"u}rst}, {Pottschmidt},
  {Caballero}, {Kreykenbohm}, {Wilms}, {Markowitz}, \&
  {Rothschild}}]{2012ApJ...745..124S}
{Suchy}, S., {F{\"u}rst}, F., {Pottschmidt}, K., {Caballero}, I.,
  {Kreykenbohm}, I., {Wilms}, J., {Markowitz}, A., \& {Rothschild}, R.~E. 2012,
  \apj, 745, 124


\bibitem[{{Torregrosa} {et~al.}(2022){Torregrosa}, {Rodes-Roca},
  {Torrej{\'o}n}, {Sanjurjo-Ferr{\'\i}n}, \&
  {Bernab{\'e}u}}]{2022RMxAA..58..355T}
{Torregrosa}, {\'A}., {Rodes-Roca}, J.~J., {Torrej{\'o}n}, J.~M.,
  {Sanjurjo-Ferr{\'\i}n}, G., \& {Bernab{\'e}u}, G. 2022, \rmxaa, 58, 355


\bibitem[{{Torrej{\'o}n} {et~al.}(2010){Torrej{\'o}n}, {Schulz}, {Nowak}, \&
  {Kallman}}]{2010ApJ...715..947T}
{Torrej{\'o}n}, J.~M., {Schulz}, N.~S., {Nowak}, M.~A., \& {Kallman}, T.~R.
  2010, \apj, 715, 947


\bibitem[{{Watanabe} {et~al.}(2003){Watanabe}, {Sako}, {Ishida}, {Ishisaki},
  {Kahn}, {Kohmura}, {Morita}, {Nagase}, {Paerels}, \&
  {Takahashi}}]{2003ApJ...597L..37W}
{Watanabe}, S., {Sako}, M., {Ishida}, M., {Ishisaki}, Y., {Kahn}, S.~M.,
  {Kohmura}, T., {Morita}, U., {Nagase}, F., {Paerels}, F., \& {Takahashi}, T.
  2003, \apjl, 597, L37


\bibitem[{{White} {et~al.}(1976){White}, {Mason}, {Huckle}, {Charles}, \&
  {Sanford}}]{1976ApJ...209L.119W}
{White}, N.~E., {Mason}, K.~O., {Huckle}, H.~E., {Charles}, P.~A., \&
  {Sanford}, P.~W. 1976, \apjl, 209, L119


\bibitem[{{Wilms} {et~al.}(2000){Wilms}, {Allen}, \&
  {McCray}}]{2000ApJ...542..914W}
{Wilms}, J., {Allen}, A., \& {McCray}, R. 2000, \apj, 542, 914


\end{thebibliography}

\end{document}